\pdfoutput=1
\documentclass[aps,prl,twocolumn,superscriptaddress,longbibliography]{revtex4-2}

\usepackage{graphicx}
\usepackage{bm}
\usepackage{hyperref}
\usepackage{amsmath}
\usepackage{amssymb}
\usepackage[table]{xcolor}
\usepackage{array}
\usepackage{multirow}
\usepackage[caption=false]{subfig}
\usepackage{physics}
\usepackage{ulem}
\usepackage[verbose]{placeins}

\usepackage{amsthm}
\usepackage{bbm}
\usepackage{multirow}
\usepackage[colorinlistoftodos]{todonotes}
\usepackage[scr=boondoxo,frak=boondox,scrscaled=1.05]{mathalfa}
\usepackage{amsmath}
\DeclareMathOperator*{\argmax}{arg\,max}

\def\tM{\tilde{M}}

\def\bra#1{\langle#1|}
\def\ket#1{|#1\rangle}
\def\braket#1#2{\langle#1|#2\rangle}

\renewcommand\vec{\boldsymbol}
\def\r{{\vec{r}}}

\def\p{{\vec{p}}}

\def\q{{\vec{q}}}
\def\A{{\vec{A}}}

\def\G{{\vec{G}}}

\newcommand\Asout{\bgroup\markoverwith{\textcolor{red}{\rule[0.5ex]{4pt}{0.6pt}}}\ULon}

\usepackage{ulem}

\makeatletter
\newcommand{\SMhidetoc}{\let\SM@l@section\l@section\let\l@section\@gobble@tw@}
\newcommand{\SMshowtoc}{\let\l@section\SM@l@section}
\makeatother

\begin{document}
\raggedbottom
\title{Shining light on collective modes in moir\'e fractional Chern insulators}

\author{Nisarga Paul}
\thanks{These authors contributed equally.}
\affiliation{Department of Physics, Massachusetts Institute of Technology, Cambridge, Massachusetts 02139, USA}
\author{Ahmed Abouelkomsan}
\thanks{These authors contributed equally.}
\affiliation{Department of Physics, Massachusetts Institute of Technology, Cambridge, Massachusetts 02139, USA}
\author{Aidan Reddy}
\thanks{These authors contributed equally.}
\affiliation{Department of Physics, Massachusetts Institute of Technology, Cambridge, Massachusetts 02139, USA}
\author{Liang Fu}
\affiliation{Department of Physics, Massachusetts Institute of Technology, Cambridge, Massachusetts 02139, USA}

\begin{abstract}
We show that collective excitations and optical responses of moir\'e fractional Chern insulators (FCIs) drastically differ from those of standard fractional quantum Hall (FQH) states in a Landau level. By constructing a variational wavefunction that incorporates the moir\'e lattice effect, we capture the collective modes in FCIs across a range of crystal momenta including the roton minimum. Interestingly, new collective modes are found in the long wavelength limit ($\bm{q} \rightarrow 0$) at low energy below the excitation continuum, distinct from the FQH case. Some of these modes are optically active and manifest as sharp peaks in optical conductivity at THz frequency. We further show that intraband optical absorption and spectral weight in twisted ${\rm MoTe}_2$ are highly tunable by the displacement field. Our work thus establishes optical spectroscopy as a powerful tool to illuminate the unique collective modes of moir\'e FCIs.
\end{abstract}

\maketitle
\addtocontents{toc}{\protect\SMhidetoc}

\textit{Introduction---} The fractional quantum anomalous Hall (FQAH) effect observed in two-dimensional moir\'e materials \cite{cai2023signatures, park2023observation, zeng2023thermodynamic, xu2023observation, Lu2024Feb}, despite occurring at zero magnetic field,  
bears a striking similarity to the fractional quantum Hall (FQH) effect in Landau levels. 
In the case of twisted transition metal dichalcogenides ($t$TMDs), 
this similarity was understood---and even anticipated \cite{devakul2021magic, li2021spontaneous, crepel2023anomalous}---by 
the close analogy between flat Chern bands \cite{wu2019topological} and Landau levels. %
The FQAH states in these partially filled Chern bands, which may be called moir\'e fractional Chern insulators (FCIs) \cite{abouelkomsan2020particle,ledwith2020fractional,repellin2020chern},  belong to the same phase as the corresponding FQH states in Landau levels. %
A natural question is whether these moir\'e FCIs host new phenomena that go beyond decades-long FQH research. %

In this work, we show that the low-energy collective modes and optical response of moir\'e FCIs are drastically different from those of FQH states.  %
For FCIs in twisted TMDs, we find that intraband collective excitations at zero wavevector $q=0$ exist below the continuum and are optically active, giving rise to sharp peaks in optical absorption at terahertz frequencies. This is in sharp contrast with FQH states in Landau levels, where %
all intra-Landau level excitations are optically dark (Kohn's theorem) \cite{Kohn1961Aug}. We further show using a sum rule that the low-energy spectral weight is highly tunable by the displacement field.   
Our work suggests low-energy optical spectroscopy as a powerful probe of collective modes  in  moir\'e FCIs.  

The starting point of our analysis is the continuum model for twisted TMD homobilayers \cite{wu2019topological}. For one valley (equivalently spin), it reads $H = H_{\uparrow} + H_C$, with $H_C$ describing Coulomb interactions and
\begin{equation}
    H_{\uparrow} = \frac{\hbar^2 \vec{k}^2}{2 m^*} \sigma_0 + \vec{J}(\vec{r}) \cdot \vec{\sigma} + V(\vec{r}) \sigma_0 \label{continuum}
\end{equation} 
where $\{\sigma_0, \vec{\sigma}\}$ are the identity and Pauli matrices in the layer-pseudospin space. $\vec{J}(\vec{r})$ is a layer Zeeman field that includes both interlayer tunneling and layer potential bias, 
and $V(\vec{r})$ %
is a periodic potential, both varying periodically with the moir\'e period. %

As we shall show below, it is conceptually useful to consider the continuum model above  in the large-$J$ limit \cite{paul2023giant, morales2023magic, zhai2020theory,shi2024adiabatic} %
under which the layer pseudospin is locally aligned to the Zeeman field, giving rise to the effective Hamiltonian $H = \tilde H + H_C$ with
\begin{equation}
\label{eq:adiabatic}
    \tilde{H} = \dfrac{(\vec{p} + e \vec A(\vec{r}))^2}{2 m^*} + \tilde{V}(\vec{r})
\end{equation} where $\vec A(\vec{r})$ is a $U(1)$ gauge field representing a periodically varying magnetic field that encloses one flux quantum per unit cell and $\tilde{V}(\vec{r}) $ is a periodic scalar potential. This adiabatic approximation maps moir\'e bands in twisted TMDs to {\it periodically modulated} Landau levels. The physical picture is that the spatially varying non-coplanar pseudospin texture gives rise to an emergent magnetic field  
$\vec{B}_e(\vec{r}) = \nabla \times \vec{A}$ that is proportional to the spin chirality \cite{bruno2004topological}.       
\par

In the following, we shall analyze neutral excitations of moir\'e FCIs using both the original model \eqref{continuum} and the adiabatic model \eqref{eq:adiabatic}, and gain a better understanding by comparing the two. In each case, we will consider a single Chern band at fractional filling, and assume throughout this work that the gap to remote bands is much larger than the interaction strength. This allows us to neglect band mixing and focus on low-energy intraband excitations.

\textit{Collective modes.--- } To study collective modes in a moir\'e FCI, we construct a variational ansatz by acting the density operator on the many-body ground state $\ket{\rm 0}$:
\begin{equation}
 \label{eq:varwavefunction}
  \ket{\phi^\alpha_{\vec{q}}} =  \sum_{\vec{G}} c^\alpha_{\vec{G}} \bar{\rho}_{\vec{q} + \vec{G}} \ket{\rm 0}
\end{equation} 
where $\vec q\in$ BZ is {\it crystal momentum}, $\vec{G}$ is a reciprocal lattice vector,  the index $\alpha$ labels different excited states at wavevector $\vec{q}$, and $\{c_{\vec G}^\alpha\}$ is a set of variational coefficients. Here, $\bar\rho_{\vec k} = P \rho_{\vec k} P$ where $\rho_{\vec k}  = \sum_i e^{-i\bm{k}\cdot\bm{r}_i}$ is the density operator at full momentum $\vec{k}$ (ranging to infinity) and $P$ is the projection operator onto the underlying Chern band subspace.

Importantly, our ansatz uses a linear combination of density operators at multiple wavevectors that differ by a reciprocal lattice vector, as allowed by the discrete lattice symmetry. This approach goes beyond the single-mode approximation ~\cite{girvin1986magneto, repellinSinglemodeApproximationFractional2014,shen2024magnetorotons}, and is crucial to capture the effect from the underlying lattice on the excited states of moir\'e FCIs, as we shall show below.   

The variational energy of Eq.~\eqref{eq:varwavefunction} is \begin{equation}\label{eq:Eqalpha}
\begin{split}
      \varepsilon_{\vec q}^\alpha&=  \dfrac{\langle \phi^\alpha_{\vec{q}} |H-E_0|\phi^\alpha_{\vec{q}}\rangle}{\langle \phi^\alpha_{\vec{q}}|\phi^\alpha_{\vec{q}} \rangle}  \\
        & = \frac{\sum_{\vec{G}_1 \vec{G}_2} c^{\alpha \ast}_{\vec{G}_1} c^\alpha_{\vec{G}_2} \langle{0}|\bar{\rho}^{\dagger}_{\vec{q} + \vec{G}_1} [H, \bar{\rho}_{\vec{q} + \vec{G}_2}] |{0}\rangle}{\sum_{\vec{G}_1\vec{G}_2} c^{\alpha\ast}_{\vec{G}_1}c^\alpha_{\vec{G}_2} \bar{s}(\vec q,\vec G_1,\vec G_2)}
\end{split}
\end{equation} 
where $\bar{s}(\vec q,\vec G_1,\vec G_2)= \langle 0|\bar{\rho}^\dagger_{\vec{q}+\vec{G}_1} \bar{\rho}_{\vec{q}+\vec{G}_2}  |0   \rangle$ is the projected static structure factor, and $(H-E_0)\ket{0} =0$. We also write $\bar s(\vec q) \equiv \bar s(\vec q,0,0)$. The variational coefficients $c^{\alpha}_{\bm{G}}$ can be found by directly diagonalizing the Hamiltonian in the subspace spanned by $\bar{\rho}_{\vec{q} + \vec{G}}\ket{0}$. %

If $\vec q$ is a high-symmetry crystal momentum, one expects strong hybridization of degenerate modes that are related by lattice symmetry. It is natural then to use the \textit{lowest star} approximation, restricting to a small set of $\vec q+\vec G$ equal in magnitude to $\vec q$. Assuming this set and the Hamiltonian have $C_n$ symmetry for some $n$, Eq.~\eqref{eq:Eqalpha} can be simplified to
\begin{equation}\label{eq:Eqell}
    \varepsilon_{\vec q}^\ell = \frac{\langle0|\bar\rho_{\vec q,\ell}^\dagger [H,\bar\rho_{\vec q,\ell}]|0\rangle}{\langle 0|\bar\rho_{\vec q,\ell}^\dagger \bar\rho_{\vec q,\ell}|0\rangle}
\end{equation}
where $\bar\rho_{\vec q,\ell} \equiv \sum_{j=0}^{n-1} e^{2\pi ij \ell/n}\bar\rho_{\vec q^{(j)}}$ is a chiral projected density operator with angular momentum $\ell$ and $\vec q^{(j)}=R_{2\pi/n}^j[\vec q]$ are the $n$-fold rotated momenta.  %

In the following, we will apply the variational ansatz to three cases: (1) ordinary FQH states, (2) the perturbed Landau level (LL) of Eq.~\eqref{eq:adiabatic}, and (3) the twisted TMD homobilayer of Eq.~\eqref{continuum}. \par 

\begin{figure}
    \centering
\includegraphics[width=\linewidth]{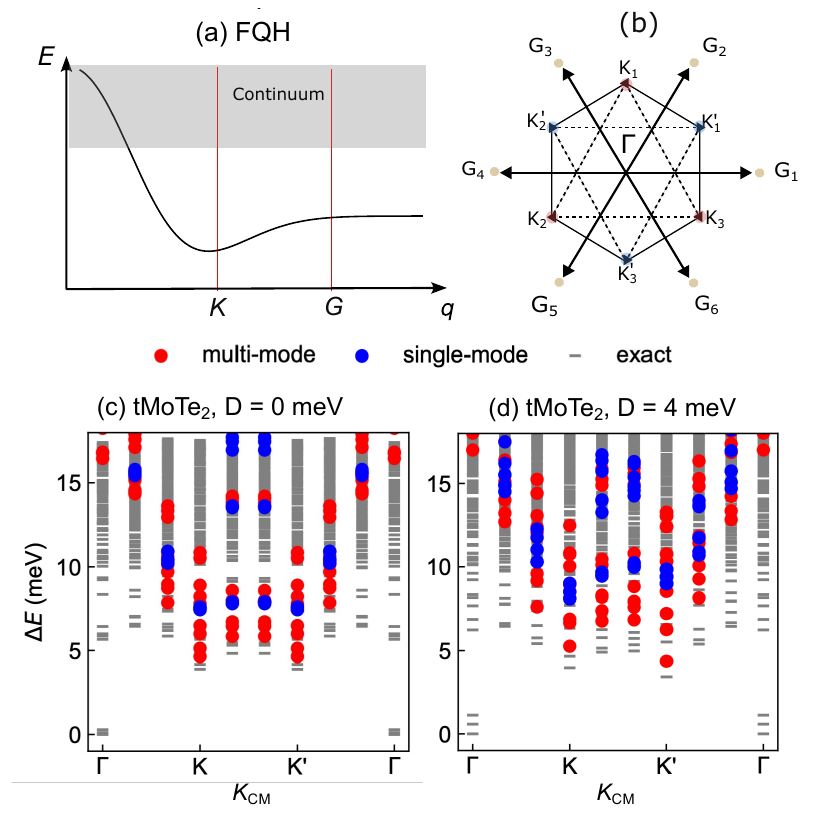}
    \caption{(a) Schematic illustration of the magneto-roton dispersion of fractional quantum Hall states at $\nu = 1/3$ (or $\nu = 2/3$). The vertical lines denote the $K, K'$ points and first shell reciprocal lattice vectors $\G$ (see panel b). (b) Illustration of the folding scheme for the $K,K'$ and $\Gamma$ points. (c-d) Many-body spectrum of twisted ${\rm MoTe}_2$ at $n= 2/3$ and twist angle $\theta = 2.5^{\circ}$ for interlayer potential bias (c) $D = 0$ meV  and (d) $D = 4$ meV. $E_0=-1027.43(-1036.96)$ meV at $D=0(4)$ meV. The red dots are variational energies from the multi-mode approximation \eqref{eq:Eqalpha}, the blue dots are from the single-mode approximation, and the grey lines are from exact diagonalization. We use a 27  unit cell cluster shown in the inset of Fig. \ref{fig:sumrule_TMDS}(b).} %
    
    \label{fig:SMA}
\end{figure}

First, we consider the standard FQH effect in the regime where the Coulomb energy $U_0 = e^2/\epsilon \ell_B$ (with $\ell_B$ the magnetic length) is small compared to the cyclotron gap $\hbar \omega_c$. In this regime, low-energy collective excitations are associated with intra-LL density fluctuations known as magneto-rotons~\cite{girvin1986magneto}. They exhibit an energy-momentum dispersion which we schematically show in Fig.~\ref{fig:SMA}(a). In this case, continuous translation symmetry reduces the variational ansatz to $\ket{\phi_{\vec q}} =\bar\rho_{\vec q}\ket{\text{FQH}}$ and Eq.~\eqref{eq:Eqalpha} to the standard single-mode approximation (SMA). %

A few features of the magneto-rotons are worth noting for our purposes. First, in the $\nu=\frac{1}{3}$ or $\frac{2}{3}$ FQH states, the magneto-roton dispersion has a minimum at a non-zero
wavevector $|\vec q_{\text{min}}|\ell_B\approx 1.5$, while it merges with the continuum %
in the long wavelength limit ($\vec{q} \rightarrow 0$) as illustrated in Fig. \ref{fig:SMA}(a). Second, while the SMA is quite accurate for small momenta up to the roton minimum, its accuracy degrades for larger momenta. Finally, the neutral excitation at large $q$ that lies below the continuum corresponds to a fractional quasielectron and quasihole separated by distance $\sim \ell_B^2 q$~\cite{laughlin1984excitons,haldane1985finite,kamilla1996excitons,Yang2012Jun,Jolicoeur2017Feb}.

Next, let us consider the case of LLs perturbed by a weak periodic potential, a special case of Eq.~\eqref{eq:adiabatic} with a uniform field. This gives the simplest model for a dispersive Chern band hosting FCIs, provided that the periodic potential is not too strong compared to the Coulomb interaction. Importantly, the periodic potential causes Bragg scattering of collective modes~\cite{kukushkin2009dispersion,wu2016moire}. To calculate the reconstructed excitation spectrum, 
we write the potential as $\tilde V(\vec r) = \sum_{\vec G} V_{\vec G}e^{-i\vec G\cdot \vec r}$, with $|\tilde V_{\vec G}|\ll \hbar \omega_c$ and project into the lowest LL. The projected Hamiltonian, including interactions, is\begin{equation}
\label{eq:LLprojected}
    \bar{H} = \sum_{\vec{G}} V_{\vec{G}}\bar{\rho}_{\vec G} + \frac{1}{2 A} \sum_{\vec{q}} U(\vec{q}) :\bar\rho_{\vec q}^\dagger \bar \rho_{\vec q}:
\end{equation}
where $\bar{\rho}_{\vec q}$ %
is the lowest LL-projected density operator %
and $U(\vec q)$ is an interaction of strength $U_0$. For concreteness, we choose a $D_6$-symmetric potential enclosing one flux quantum per unit cell with six harmonics: $\tilde V_{\vec{G}} = - V_0$ for $\vec{G}_i= \frac{4 \pi}{\sqrt{3} a_0} [\cos 2\pi (i-1)/6,\sin 2\pi (i-1)/6]$ for $ i = 1,\dots,6$ where $\sqrt{3}a_0^2=4\pi\ell_B^2$.
\par 

The Bragg scattering of magneto-roton modes by the periodic potential is %
captured by our variational ansatz Eq.~\eqref{eq:varwavefunction}, which is a linear combination of excited states at wavevectors $\vec{q} + \vec{G}$ as shown by the momentum folding scheme in Fig.~\ref{fig:SMA}(b). This can be viewed as a natural generalization of the SMA from FQH liquids to FCIs with discrete translation symmetry.  
In the Supplemental Material~\cite{supp_arXiv}, we show how the variational energy can be expressed in terms of the structure factor and ground-state expectation value of $\bar\rho_{\vec G}$. 
\par 
Furthermore, if $V_0\ll U_0$, the variational energy of excited states can be directly expressed in terms of FQH ground-state expectation values. For instance, to leading order in $V_0$, the energies of the modes at the $K$ point of the Brillouin zone (BZ), which we refer to as $K$-rotons, take the form 
\begin{equation}\label{eq:Krotonenergy}
    \varepsilon_{K}^\ell =  \Delta_K -\frac{|V_0|}{\bar s_0(K)} ( S_3 e^{i 2\pi \ell/3} + c.c.)
\end{equation}
where $\ell$ labels $C_3$-angular momentum, $\Delta_K$ is the energy of the FQH magneto-roton at wavevector $K$, and $S_3$ is three-density correlator defined as:  
\begin{equation}\label{eq:3rho}
    S_3 = \langle \bar\rho_{-K_1} \bar\rho_{K_1-K_2} \bar \rho_{K_2} \rangle, 
\end{equation}
where the expectation value is taken in the FQH ground state in the absence of periodic potential~\cite{supp_arXiv}.  
We note that $S_3$ could in principle be evaluated for a model wavefunction (such as Laughlin) using Monte Carlo~\cite{Morf1986Feb}. 

Finally, we turn to the moir\'e system twisted MoTe$_2$ ($t$MoTe$_2$),  which hosts an FCI with threefold rotational symmetry. %
Indeed, in the adiabatic limit of Eq.~\eqref{eq:adiabatic}, this model maps onto the perturbed Landau level model with the periodic potential enclosing one flux quantum per unit cell. %
In contrast to the previous case, the variational energies must be evaluated numerically. We perform band-projected exact diagonalization (at hole filling $n=2/3$ of the $C=1$ miniband at twist angle $\theta=2.5^\circ$ and displacement field $D$~\cite{Reddy2023Aug0}), 
and compare the results with variational energies obtained by diagonalizing the Hamiltonian in the span of all states $\bar{\rho}_{\bm{q}+\bm{G}}\ket{\text{GS}_i}$ satisfying $|\bm{q}+\bm{G}|\leq \frac{4\pi}{\sqrt{3}a_0}$ where $i$ labels the three quasi-degenerate ground states.

We find that the variational ansatz captures the structure of the low-lying excitations near the $K$ and $K'$ points as shown in Fig. \ref{fig:SMA}(c,d). At zero displacement field $D=0$, $C_{2y}\mathcal{T}$ symmetry relates the $K$ and $K'$ moiré center-of-mass crystal momentum sectors in a fixed valley, enforcing degeneracy between these sectors' many-body spectra.  A nonzero $D$ breaks $C_{2y}$ symmetry, thus lifting the $K/K'$ degeneracy as apparent in Fig. \ref{fig:SMA}(d). In particular, the roton minimum softens more strongly at $K'$ than $K$. Our variational approach succeeds in reproducing this trend.
\par 

The success of our variational ansatz near $K,K'$ in $t$MoTe$_2$ can be qualitatively understood by the adiabatic mapping to perturbed FQH.  As illustrated in Fig. \ref{fig:SMA}(a),
the $K$ and $K'$-rotons are derived from magneto-roton modes with $q\ell_B = |K|\ell_B =\frac{2 \sqrt{\pi }}{3^{3/4}}\approx 1.6$, which is close to the magneto-roton minimum of the unperturbed FQH liquids (Fig.\ref{fig:SMA}(a)) where the SMA is quite accurate. Moreover, Bragg scattering hybridizes magneto-roton modes at $K_1$, $K_2$ and $K_3$, giving rise to three non-degenerate roton modes at the $K$ point in moir\'e BZ, and likewise for $K'$. In Fig. \ref{fig:SMA}(c,d) the multi-mode approximation is more accurate than the single-mode approximation, showing that the Bragg scattering is quantitatively significant.

In contrast with the $K$-rotons, the collective modes at $\Gamma$ %
are poorly captured by the variational ansatz. Figs. \ref{fig:SMA}(c,d) show that the lowest excited states from exact diagonalization at $\Gamma$ are isolated from the continuum and considerably lower in energy than the variational states.
This discrepancy is once again consistent with the mapping to FQH, see Fig. \ref{fig:SMA}(a); variational states at $\Gamma$ are linear combinations of modes at $\vec{G}_i$ with $|\vec G_i|\ell_B\approx 2.7$, where the SMA breaks down and collective excitations are better described as quasielectron-quasihole pairs.%

Based on the mapping to perturbed FQH, we expect that the low-energy neutral excitations of $t$TMD at $\Gamma$ shown in Fig. \ref{fig:SMA}(c,d) are quasielectron-quasihole pairs, analogous to their FQH counterpart at momentum $\vec{G}$ (which folds back to $\Gamma$).      
A particularly interesting possibility is that these quasielectron-quasihole pairs may exist at energies {\it below} the excitation continuum. Indeed, this scenario is expected to occur when our system maps to {\it weakly} perturbed FQH and is consistent with the exact diagonalization data shown in Figs. ~\ref{fig:SMA}(c,d), as we elaborate in the Supplemental Material \cite{supp_arXiv}. %
The nature of low-energy neutral excitations at $\Gamma$ is of great interest as they may affect the optical response of the system $\sigma(\vec{q} \rightarrow 0, \omega)$. 
Although our variational ansatz based on SMA becomes inaccurate here, alternative approaches (e.g., based on composite fermion wavefunctions) \cite{kamilla1996excitons, Yang2012Jun} may better capture the quasielectron-quasihole pairs. 

\textit{Low-frequency optical absorption.---} It is well-known that the intra-LL excitations of FQH states have a vanishing dipole coupling to light in the absence of disorder~\cite{Kohn1961Aug}. This is because a uniform electric field only couples to electrons' center of mass, which decouples from other degrees of freedom and has zero intra-LL matrix elements. As a result, optical absorption occurs  only at the cyclotron frequency, leaving low-energy collective modes of FQH states optically dark. 

In contrast, moir\'e FCI states only have discrete translation symmetry to which Kohn's theorem does not apply. Thus, a direct coupling between light and $\vec q=0$ collective excitations with angular momentum $\ell=\pm 1$ is allowed by symmetry. In particular, low-energy quasielectron-quasihole pairs may be isolated from the continuum and thus undamped,   %
raising an interesting question of whether they are optically active and can be detected as sharp peaks in the optical conductivity at low frequency.\par  

Indeed, our calculations of the optical response to circularly polarized light in the many-body ground state of twisted ${\rm MoTe}_2$ reveal sharp and prominent peaks at frequencies that correspond to low-lying chiral modes ($\ell=\pm 1$) at $\Gamma$, which are further enhanced by displacement fields \cite{supp_arXiv}.

To better understand the low-frequency optical response, we derive an intraband $f$-sum rule for the low-energy optical spectral weight for an FCI formed in an isolated Chern band separated from remote bands by a large gap $\Delta$. 
In particular, we relate the low-energy spectral weight to a simple ground-state expectation value. 

To derive this sum rule, we first perform a Schrieffer-Wolff transformation---a {\it unitary} transformation $U$---to decouple the low-energy and high-energy subspace by eliminating the off-diagonal matrix elements (i.e., interband terms): 
\begin{eqnarray}
\tilde{H} = U H U^{-1} = \tilde{H}_1 + \tilde{H}_2
\end{eqnarray}
where $P \tilde{H}_1 P =\tilde{H}_1$ is the low-energy Hamiltonian acting on a low-energy subspace specified by the projection operator $P$, and $P_{\perp} \tilde{H}_2 P_{\perp} = \tilde{H}_2$ with $P_{\perp}=I-P$ associated with the high-energy subspace.    

To compute the optical conductivity, we find it convenient to relate it to the dynamical structure factor through the continuity equation and fluctuation-dissipation theorem (a thorough discussion for Coulomb systems can be found in Ref.~\cite{onishi2024quantum}): 
\begin{equation}\label{eq:Stosigma}
        S(\bm{q},\omega) = \frac{\hbar A}{\pi e^2}\frac{q_aq_b}{\omega}\Re \sigma_{ab}(\bm{q},\omega),
\end{equation}
where $A$ is the system area, and the dynamical structure factor is defined as
\begin{eqnarray}\label{eq:dsf}
    S(\bm{q},\omega) &=& \sum_{n\neq 0}\delta\left(\omega-\omega_{n0}\right)|\bra{n}\rho_{\bm{q}}\ket{0}|^2  \nonumber \\
    &=& \sum_{n\neq 0}\delta\left(\omega-\omega_{n0}\right)|\bra{\tilde{n}} \tilde{\rho}_{\bm{q}}\ket{\tilde{0}}|^2. 
\end{eqnarray}
Here $n$ labels the many-body eigenstate with energy $\hbar\omega_n$, $\omega_{n0}=\omega_n-\omega_0$, and $\rho_{\bm{q}}$ is the density operator. 
In the second equality, we have rewritten $S$ in terms of the many-body eigenstates $\ket{\tilde{n}} \equiv U \ket{n}$ for the transformed Hamiltonian $\tilde{H}$, and $\tilde{\rho}_{\bm{q}} = U\rho_{\bm{q}} U^{-1}$. 

As $\tilde{H}$ decomposes into low- and high-energy parts, the structure factor is naturally composed of an intraband part that only involves low-energy eigenstates denoted as $\ket{\tilde l}$ and an interband part that involves matrix elements between the ground state and high-energy eigenstates denoted as $\ket{\tilde h}$, 
\begin{eqnarray}
S(\bm{q},\omega) &=& \sum_{l\neq 0}\delta\left(\omega-\omega_{l0}\right)|\bra{\tilde{l}} \tilde{\rho}_{\bm{q}}  \ket{\tilde{0}}|^2  \nonumber \\
&+&  
\sum_{h}\delta\left(\omega-\omega_{h0}\right)|\bra{\tilde{h}}  \tilde{\rho}_{\bm{q}}  \ket{\tilde{0}}|^2 \nonumber \\
&\equiv& S^L(\vec{q}, \omega) +  S^H(\vec{q}, \omega), 
\end{eqnarray}
with $P\ket{\tilde{l}}=\ket{\tilde{l}}$ and $P_\perp \ket{\tilde{h}}=\ket{\tilde{h}}$. 
Following Eq.~\eqref{eq:Stosigma}, the optical conductivity can be similarly decomposed: $\sigma=\sigma^L + \sigma^H$. In particular, we shall focus on the structure factor and optical conductivity at low frequency, which are dominated by the intraband contributions as captured by the low-energy Hamiltonian $\tilde{H}_1$ \cite{mao2023diamagnetic}. 

We define the intraband optical spectral weight as
\begin{equation}\label{eq:F}
    W = \int_0^{\infty} d\omega\, \Re\sigma^L_{aa}(\omega).
\end{equation}
From Eq.~\eqref{eq:Stosigma} it follows that 
\begin{equation}
\label{eq:spectralweight}
\begin{split}
    W &= \frac{\pi e^2}{\hbar A} \lim_{ q\to 0} \frac{1}{q^2}\sum_{l \neq 0}  |\bra{\tilde{l}}\tilde{\rho}_{\bm{q}}\ket{\tilde{0}}|^2  \omega_{l0}\\
    &= \frac{\pi e^2}{2\hbar^2 A} \lim_{q\to 0} \frac{1}{q^2} \langle \tilde{0} | [ \tilde{\rho}_{\vec q}^{L \dagger},[\tilde{H}_1, \tilde{\rho}^L_{\vec q} ]]| \tilde{0}\rangle
\end{split}
\end{equation}
where $\tilde{\rho}^L \equiv P \tilde{\rho} P$ is the density operator projected onto the low-energy subspace. 

For our system, the isolated Chern band defines a low-energy subspace, separated by a band gap $\Delta$ to remote bands. The corresponding Schrieffer-Wolff transformation can be carried out perturbatively in $\lambda/\Delta$: $U=I + \mathcal{O}(\lambda/\Delta)$, where $\lambda H' \equiv P H P_{\perp} + P_{\perp} H P$ is the interband part of  Coulomb interaction with interaction strength $\lambda$. To the zeroth-order approximation in $\lambda/\Delta$, we have $U\approx I$ and $\ket{\tilde{0}}\approx \ket{0}$ is the ground state of band-projected Hamiltonian $\bar{H} \equiv P H P \approx \tilde{H}_1$.  Then, the low-energy optical spectral weight simplifies to 
\begin{equation}
\label{eq:spectralweight-band}
\begin{split}
    W = %
    \frac{\pi e^2 n}{\hbar^2}\lim_{q\to 0} \frac{1}
    {q^2} F(\q) 
\end{split}
\end{equation} with \begin{equation}
\label{eq:oscillator_q}
    F(\q) \equiv \dfrac{1}{2 N} \langle 0 | [ \bar{\rho}_{\vec q}^{\dagger},[\bar{H}, \bar{\rho}_{\vec q} ]]| 0\rangle
\end{equation}
where $\bar{H}$ and $\bar{\rho}$ are the band-projected Hamiltonian and density operator respectively. Alternatively, 
the right-hand side of Eq.~\eqref{eq:spectralweight-band} can be expressed using electron position operator~\cite{mao2023diamagnetic}, which is well defined for systems with open boundary conditions. 

\par As before, we apply this to the three cases of (1) ordinary FQH states, (2) perturbed Landau levels, and (3) twisted TMD homobilayers. In the case of ordinary Landau levels, only the projected interactions $H_C = \frac{1}{2A} \sum_{\vec q}U(\vec q)\bar\rho_{\vec q}^\dagger \bar\rho_{\vec q}$ remain. We may write
\begin{align}
    F(\vec{q}) = \frac{1}{2A} \sum_{\vec q'} U(\vec q') f_{\vec q'}^{\vec q} \left(f_{\vec q}^{\vec q'} \bar s(\vec q+\vec q') + f_{-\vec q}^{\vec q+\vec q'} \bar s(\vec q')\right)\nonumber
\end{align}
using the GMP algebra $[\bar\rho_{\vec q},\bar\rho_{\vec q'}] = f_{\vec q}^{\vec q'}\bar \rho_{\vec q+\vec q'}$ where $f_{\vec q}^{\vec q'} = 2ie^{\vec q\cdot \vec q'\ell_B^2/2}\sin(\vec q\wedge \vec q' \ell_B^2/2)$. Using parity symmetry, it follows $F(\vec q)$ vanishes to order $q^4$~\cite{girvin1986magneto,supp_arXiv}, and hence $W=0$ in this case, which is also consistent with Kohn's theorem. \par 

In the case of Landau levels perturbed by a periodic potential, the projected Hamiltonian is $\bar H = \sum_{\vec G} V_{\vec G} \bar\rho_{\vec G} + H_C$. The contribution from $H_C$ vanishes as above, and employing the GMP algebra then yields~\cite{wu2016moire,mao2024low}
\begin{equation}\label{eq:WpertLL}
    W = \frac{\pi e^2}{2\hbar^2 A}\sum_{\vec G}  \left(- V_{\vec G} \cdot \rho(\G)\right)(\hat q \wedge \vec G\ell_B^2)^2 
\end{equation}
where $\rho(\G) \equiv \langle 0|\bar\rho_{\vec G}|0\rangle = \langle 0|\rho_{\vec G}|0\rangle$ is the Fourier component of the charge density in the {\it exact} ground state $\ket{0}$ of $\bar H$. It is important to note that $\rho(\G)$ is generally nonzero because of the periodic potential, leading to nonzero intraband spectral weight $W>0$. Hence the perturbed Landau level serves as perhaps the simplest example of an FCI with nonzero optical spectral weight.

\begin{figure}
    \centering        
    \includegraphics[width=\linewidth]{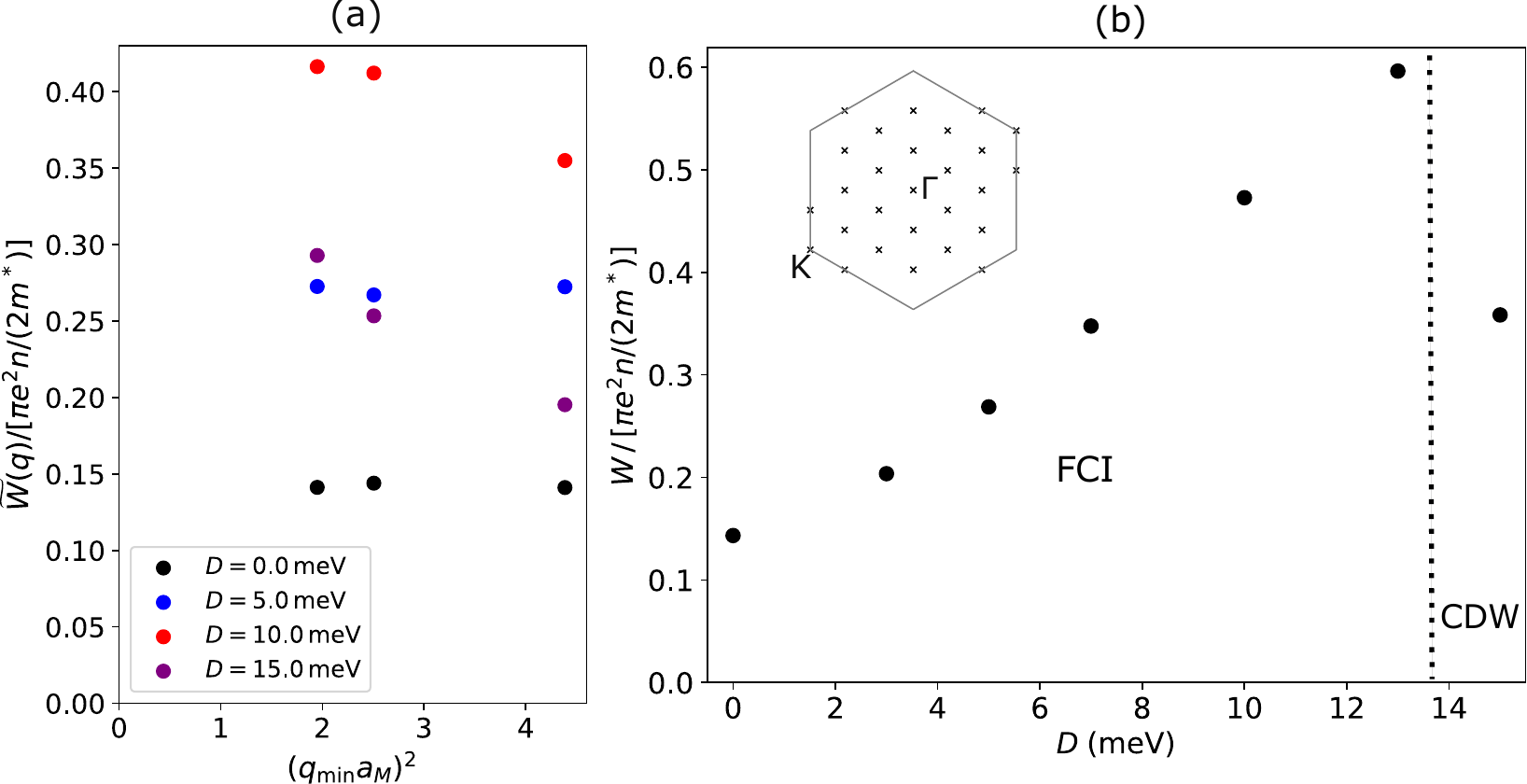}  
    \caption{(a) $\widetilde{W}(\q) = (\pi e^2 n F(\q))/(\hbar^2 q^2) $ evaluated at the smallest allowed non-zero momentum point $q_{\rm min}$ for the finite systems we considered (with  12, 21 and 27 unit cells) for various displacement fields $D$. (b) Intraband optical weight $W = \lim_{q \rightarrow 0} \widetilde{W}(\q) $, calculated by extrapolating  $\widetilde{W}(\q_{\rm min})$ with a linear fit, as a function of displacement field $D$. Calculations are done for twisted ${\rm MoTe}_2$  at $n= 2/3$ and twist angle $\theta = 2.5 ^\circ$. $W$ is given in units of $\pi e^2 n/(2m^*)$ where $n$ is the density and $m^*=0.62 m_e$ is the effective mass. Inset: 27 unit cell cluster, the largest of the three used.}      \label{fig:sumrule_TMDS}
\end{figure}

For a generic Chern band such as the bands of twisted ${\rm MoTe}_2$, it is difficult to compute $W$ analytically. In the Supplemental Material, we use Eq. \ref{eq:WpertLL} to evaluate the optical spectral weight of the adiabatic model for $t$MoTe$_2$ projected to the lowest Landau level \cite{supp_arXiv}. %
Without these approximations, we can still evaluate $W$ for $t$MoTe$_2$ using the many-body ground state obtained from band-projected exact diagonalization. In Fig. \ref{fig:sumrule_TMDS}(a), we compute $\widetilde{W}(\q) = (\pi e^2 n F(\q))/(\hbar^2 q^2)$ evaluated at the smallest allowed non-zero momentum $q_{\min}$ for the finite-size systems we studied. As clearly shown, $\widetilde{W}$ extrapolates to a finite value $W = \lim_{\q \rightarrow 0} \widetilde{W}(q)$ for various displacement fields. The finiteness of $W$ extends to a wide range of twist angles \cite{supp_arXiv}, demonstrating the robust existence of low-frequency optical absorption, in contrast to recent studies \cite{shen2024magnetorotons,wolf2024intraband}. 

We find that the low-frequency optical absorption is highly tunable with displacement field $D$ (Fig. \ref{fig:sumrule_TMDS}(b)). As $D$ increases, we observe increasing spectral weight in the FCI phase, consistent with the reduced similarity between the underlying Chern band and the lowest Landau level. We also observe an abrupt decrease in spectral weight as the system undergoes a phase transition to a charge density wave at large $D$~\cite{Sharma2024Sep,zaklama2024structure}. The optical weight reaches an $O(1)$ fraction of the sum-rule limit $\pi e^2n/2m^*$, supporting the experimental visibility of the optical response.
\par 

\textit{Discussion.---} %
Our findings suggest new directions for optical experiments on moir\'e FCIs. The {\it chiral} %
fractional quasielectron-quasihole excitations of $t$TMDs, with angular momentum $\ell=\pm 1$ and few-meV excitation energy, can be probed directly using  terahertz (THz) spectroscopy, or indirectly using inelastic Raman scattering (note that $\ell = \pm 1 = \mp 2 \mod 3$). We predict that the intraband optical absorption increases with $D$ field, which can either be observed directly at low frequencies or indirectly at higher frequencies as a \textit{decrease} in the interband optical absorption with $D$. 
\par
The predicted optical response lies at energies of a few meV, corresponding to the THz regime. This scale is challenging but increasingly accessible in van der Waals devices including TMD semiconductors, where recent THz measurements have reached frequency windows of 0.1-1 THz and allow electrostatic control by gates~\cite{seo2024chip,chen2025terahertz,pierce2025imaging}. Since the intraband optical spectral weight can reach $0.1$--$0.6$ of the $f$-sum rule limit in the FCI phase, collective mode optical absorption is not expected to be parametrically weak in comparison to interband absorption in twisted TMD homobilayers~\cite{wolf2024intraband}. 

\par 
In addition to magnetorotons, FQH states also host chiral graviton modes at small $\vec{q}$ \cite{haldane_geometrical_2011,liou_chiral_2019,yang2016acoustic}, which can be probed by Raman scattering \cite{liang2024evidence}. We emphasize that such graviton modes \cite{shen2024magnetorotons,wang2025dynamics,long2024spectra} will not be the lowest energy excited states in twisted ${\rm MoTe}_2$ due to the existence of quasielectron-quasihole excitations that are lower in energy. In the Supplemental Material \cite{supp_arXiv}, we provide numerical evidence that chiral gravitons in periodically modulated LLs are optically active, and leave its detailed study to the future. 
\par 
While we have focused on the spin-polarized FCIs at $\nu=1/3$ and $\nu=2/3$, our variational approach carries over to the FCI analogs of other Jain sequence states, even-denominator non-Abelian FCIs~\cite{Reddy2024Oct,Xu2024Mar,Ahn2024Oct,wang2024higher},  non-Abelian spin Hall insulators at $\nu=3$~\cite{Abouelkomsan2024Jun}, and potentially composite Fermi liquids~\cite{goldman2023zero,dong2023composite,abouelkomsan2024compressible,stern_transport_2023}. Based on our intraband optical sum rule Eq. ~\ref{eq:spectralweight-band}, we expect %
nonzero intraband optical absorption to occur in all of these states as well. Other interesting future directions include developing variational descriptions of the optically active intraband excitations, for example using composite-fermion-based wave functions~\cite{kamilla1996excitons,Hu2024Jun}, as well as the condensation of $K$-rotons leading to a fractional quantum anomalous Hall crystal~\cite{song2024intertwined}, Hall viscosity in FCIs~\cite{Haldane2009Jun,Read2011Aug}, and extensions to lattice FCIs and higher topological bands. 
\par 

\begin{acknowledgments}
\textit{Acknowledgments.--- }  We thank Yugo Onishi for related collaborations, and Kun Yang, Girsh Blumberg, Debanjan Chowdhury, Ady Stern, and Long Ju for useful discussions. This work was supported by the Air Force Office of Scientific Research (AFOSR) under
 Award No. FA9550-22-1-0432.  The authors acknowledge the MIT SuperCloud and Lincoln Laboratory Supercomputing Center for providing computing resources that have contributed to the research results reported within this paper. A.A. was supported by the Knut and Alice Wallenberg Foundation (KAW2022.0348). L.F. was partly supported by the Simons Investigator Award from the Simons Foundation.
\end{acknowledgments}

\bibliography{ref}

\clearpage
\onecolumngrid
\begin{center}
\textbf{\large Supplemental Material}
\end{center}

\setcounter{secnumdepth}{4}
\setcounter{section}{0}
\setcounter{equation}{0}
\setcounter{figure}{0}
\setcounter{table}{0}
\renewcommand{\thesection}{S\arabic{section}}
\renewcommand{\theequation}{S\arabic{equation}}
\renewcommand{\thefigure}{S\arabic{figure}}
\renewcommand{\thetable}{S\arabic{table}}
\renewcommand{\theHsection}{\thesection}
\renewcommand{\theHequation}{\theequation}
\renewcommand{\theHfigure}{\thefigure}
\renewcommand{\theHtable}{\thetable}

\addtocontents{toc}{\protect\SMshowtoc}
\tableofcontents

\section{Roton folding in periodic systems and finite-size effects}\label{sec:SM_folding}

In this section, we discuss the folding of the roton spectrum in crystalline systems and point out a finite-size effect for the counting of expected folded roton modes based on symmetry considerations. We consider the Hamiltonian of the LLL which takes the form, 

\begin{equation}
\label{eq:supp_LL}
    H = \frac12 \sum_{\vec{q}} U(\vec{q}) :\bar\rho_{\vec q}^\dagger \bar \rho_{\vec q}:
\end{equation}
where $\bar{\rho}_{\vec q} = \sum_je^{-|\vec q|^2\ell_B^2/4} e^{-i\vec q\cdot \vec R_{j}}$ is the LL-projected density operator in the $n=0$ LL in terms of particle guiding center coordinates $\vec R_j$. The interaction $U(\vec{q})$ is taken to be bare Coulomb $U(\vec{q}) = \frac{U_0 a_0}{2 A}  \frac{2 \pi}{|\vec{q}|}$. with the interaction strength $U_0 = e^2 /( 4 \pi \epsilon a_0)$.

 As shown by Haldane \cite{haldane1985many}, many-body magnetic translations in this case are factorized into a relative and a center of mass (COM) part. The momentum eigenvalues corresponding to the relative motion can be used to label any many-body state. On a finite torus spanned by two vectors $\vec{L}_1$ and $\vec{L}_2$ such that $|\vec{L}_1 \cross \vec{L}_2| = 2 \pi \ell_B^2 N_\phi$, the relative momentum takes discrete values, \begin{equation}
 \label{eq:FQHBZ}
     \vec{q} \in {\rm BZ_{cont}} = m \vec{g}_1 + n \vec{g}_2  \> \> , \> \>  (m,n) = (0 \dots N-1, 0 \dots N -1)
 \end{equation} where $N = {\rm GCD}(N_p,N_\phi)$ with ${\rm GCD}$ denoting greatest common divisor, $N_p$ and $N_\phi$ are the number of particles and flux quanta respectively. $\vec{g}_1$ and $\vec{g}_2$ are two reciprocal lattice vectors such that $\vec{L}_i \cdot \vec{g}_j = 2 \pi \delta_{ij}$. The discretization defines a \textit{many-body} Brillouin zone which we will refer to as ${\rm BZ_{cont}}$. As system size increases (while keeping a fixed filling $\nu$), the mesh defined by $\vec{q}$ becomes denser and larger and in the thermodynamic limit, $\vec{q} \in R^2$ as expected from a continuous system. In addition to the relative motion, COM motion gives rise to $q$-fold degeneracy for $\nu = p/q$ such any state at momentum $\vec{q} = (q_1,q_2)$ is degenerate with other states at $\vec{q} = (q_1, q_2 + m N)$ for $m = 1, \cdots q-1$.

\begin{figure}[t!]
    \centering
    \includegraphics[width=\linewidth]{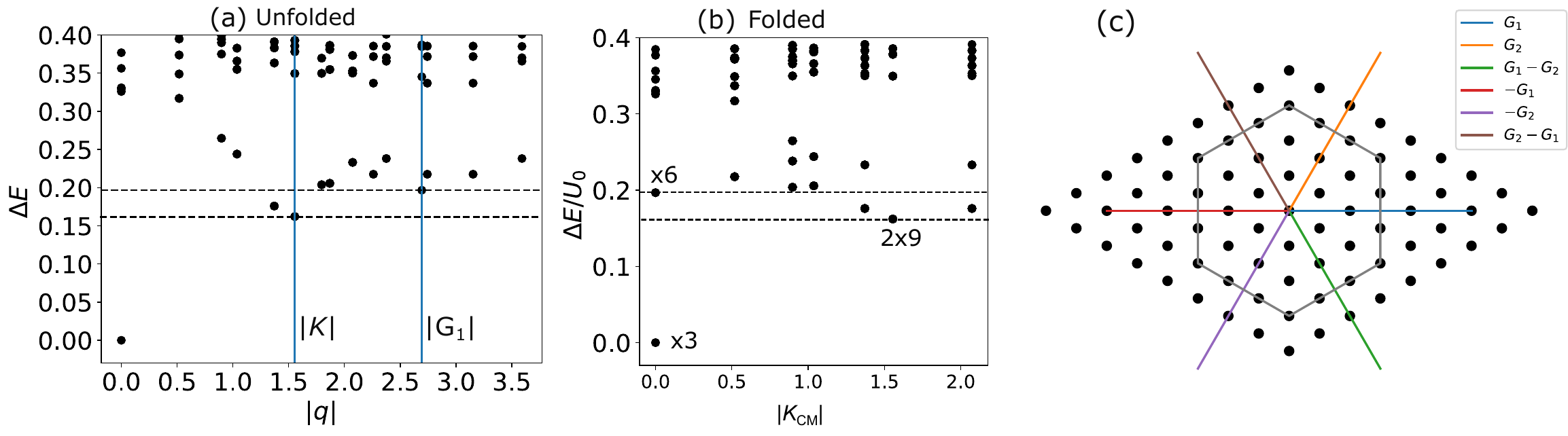}
    \caption{(a) The unfolded $\nu = 1/3$ fractional quantum Hall many-body spectrum as a function of $|\vec{q}| \in {\rm BZ_{cont}}$ (equation \eqref{eq:FQHBZ}) obtained on a hexagonal torus with $N_\phi = 27$ and equal aspect ratio. The vertical lines denote the magnitude of the momentum points $\vec{K}$ (corner points of ${\rm BZ_{Bloch}}$) and $\vec{G}_1$ (reciprocal lattice vector of ${\rm BZ_{Bloch}}$) (c.f. panel (c)). (b) The folded $\nu = 1/3$ many-body spectrum on the same finite system plotted as a function of $|\vec{K}_{\rm CM}| \in {\rm BZ_{Bloch}}$ (equation \eqref{eq:BlochBZ}) obtained from working with magnetic Bloch wavefunction for the LLL. (c) Illustration of ${\rm BZ_{cont}}$ (all black points) and $\rm BZ_{Bloch}$ (all black points bounded by the hexagon) for a hexagonal torus with $N_1 \times N_2 = N_\phi = 27$. One immediately sees that not all first shell $\vec{G}$ lie in ${\rm BZ_{cont}}$. Only $\vec{G}_1$ and $-\vec{G}_1$ lie in ${\rm BZ_{cont}}$ which means only two states (for each topologically distinct ground state) fold back to $\vec{K}_{\rm CM} = 0$ giving rise to a total of six states as shown in panel (b). }
    \label{figS1}
\end{figure}

In Fig \ref{figS1}(a), we show the many-body spectrum obtained from diagonalizing equation \eqref{eq:supp_LL} utilizing the many-body quantum numbers $|\vec{q}|$  at $\nu = 1/3$ (equivalently $\nu = 2/3$). We use a hexagonal torus with $N_\phi = 27$ and equal aspect ratio. We observe as expected the FQH ground state at $|\vec{q}| = 0$. In addition, the magneto-roton branch is clearly visible. Notice that the magneto-roton minima on this cluster occur at the $K$ points, defined as $|K| \ell_B \approx 1.6$.

To make connection with periodic crystalline system with a well-defined unit cell and discrete translation symmetry, it's natural to choose the single particle LLL wavefunctions to be magnetic Bloch basis $\phi_{\vec{k}}(\vec{r})$ parametrized by a Bloch momentum $\vec{k}$ \cite{wang2021exact} that lives in a Brillouin zone which we will refer to as ${\rm BZ_{Bloch}}$. This is obtained by defining a unit cell spanned by two vectors $\vec{a}_1$, $\vec{a}_2$ that encloses one flux quantum $|\vec{a}_1 \cross \vec{a}_2| = 2 \pi \ell_B^2$ and applying Bloch's theorem. Again, on a finite torus spanned by two vectors $\vec{L}_1 = N_1 \vec{a}_1$, $\vec{L}_2 = N_2 \vec{a}_2$ with $N_1 \times N_2 = N_\phi$, we have
 \begin{equation}
 \label{eq:BlochBZ}
    \mathbf{k} \in {\rm BZ_{Bloch}} = m \vec{g}_1 + n \vec{g}_2  \> \> , \> \>  (m,n) = (0 \dots N_1 -1, 0 \dots N_2 -1)
\end{equation} where $\vec{L}_i \cdot \vec{g}_j = 2 \pi \delta_{ij}$. 
The many-body Hamiltonian diagonalized with these quantum numbers \textit{does not} reveal the full many-body symmetries of the problem. In such a case, the total momentum $\vec{K}_{\rm CM} = \sum_i^{N_p} \vec{k}_i$ is \textit{folded} relative to the relative momenta $\vec{q} \in {\rm BZ_{cont}}$  as $\vec{k}_i$ is defined only modulo reciprocal lattice vectors $\vec{G} = m \vec{G}_1 + n \vec{G}_2 $ with $\vec{G}_i \cdot \vec{a}_j = 2\pi \delta_{ij}$  \cite{bernevig2012emergent,repellinSinglemodeApproximationFractional2014}. 

To illustrate this, we work again with the same hexagonal torus that corresponds to the many-body spectrum in Fig. \ref{figS1}(a). We diagonalize the Hamiltonian \eqref{eq:supp_LL} as a function of the Bloch momentum $|\vec{K}_{\rm CM}|$. As shown in Fig \ref{figS1}(b), the many-body energy states are folded relative to what is plotted in Fig. \ref{figS1}(a). In addition, the many-body spectrum now contains multiple degeneracies from the COM motion that have been factored out in the spectrum shown Fig. \ref{figS1}(a). 

To study the low-lying excitations of crystalline systems which do not exhibit continuous translational symmetry and have momentum only defined modulo $\vec{G}$, it's crucial to understand the number of states that fold back to a generic many-body Bloch momenta $K_{\rm CM} \in {\rm BZ_{Bloch}}$ from a momentum point $\vec{q} \in {\rm BZ_{cont}}$. Since $K_{\rm CM}$ is defined only modulo $\vec{G}$, all states at momentum $\vec{q} = K_{\rm CM} + \vec{G}$ will map back to $K_{\rm CM}$. 

As we discussed earlier, the momentum $\vec{q} \in {\rm BZ_{cont}}$ is discretized on a finite torus. While all $\vec{G}$ are equivalent in a crystalline system with discrete translational symmetry, they are physically distinct momenta in  the ${\rm BZ_{cont}}$. \textit{Crucially}, on a finite system,  ${\rm BZ_{cont}}$ will not contain all distinct $\vec{G}$ in it. It's only in thermodynamic limit that ${\rm BZ_{cont}} \rightarrow R^2$ and therefore all $\vec{G}$ are covered. Also, as a finite-size effect, not all $\vec{G}$ related by rotation can exist in the ${\rm BZ_{cont}}$ in finite system. In our example (Fig. \ref{figS1}(b)), naively one expects (based on the folding picture) a low-lying set of $3 \times 6 = 18$ states at $\vec{K}_{\rm CM} \equiv \Gamma  = 0$ above the ground state. These are the rotons that fold from the first shell $\vec{q} = m \vec{G}_1 + n \vec{G}_2$ for $(m,n) = (1,0),(0,1),(1,-1),(-1,1),(-1,0),(0,-1) $ which are related by $C_6$ rotation. The factor of 3 corresponds to the three topologically degenerate ground states. However, we only observe 6 states at $K_{\rm CM} = \Gamma$ (Fig. \ref{figS1}(b)). To understand the origin of these, we plot in Fig. \ref{figS1}(c) the distinct $\vec{q} \in {\rm BZ_{cont}}$ and the distinct $K_{\rm CM} \in {\rm BZ_{Bloch}}$ for the finite system we work with. We observe that ${\rm BZ_{FQH}}$ can only fit two $\vec{G}$ vectors out of the first shell. This gives rise to two states for each topologically distinct ground state leading to total of six states which what we observe in Fig. \ref{figS1}(b) at $K_{\rm CM} = \Gamma$. 

Unfortunately, this finite-size effect exists for all numerically accessible finite-size systems which poses a challenge in studying the properties of the %
quasielectron-quasihole states at $\Gamma$ as we only observe a subset of the states that fold from the first shell $\vec{G}$ vectors. On the other hand, the situation is less drastic for the $K$ and $K'$ rotons. Based on the folding picture, we expect three low-lying states  for each ground state corresponding to three equivalent $K$ or $K'$ point giving rise to a total of 9 states. Indeed, this is what we observe in Fig. \ref{figS1}(b).

In all cases, on all accessible finite size systems, we cannot access states that fold from momentum points beyond the first shell of $\vec{G}$ vectors. However, these states will be generally higher in energy and most likely won't contribute to the low energy properties of the system. 

\section{Variational ansatz in periodically modulated Landau levels}\label{sec:SM_variational}

In this section, we detail the application of the variational ansatz to the periodically modulated lowest Landau level (LLL). We study the LLL perturbed by a weak periodic potential enclosing one flux quantum per unit cell. We recall the projected Hamiltonian \begin{equation}
    H = \sum_{\vec{G}}  V_{\vec{G}}\bar{\rho}_{\vec G} + \frac{1}{2A} \sum_{\vec{q}} U(\vec{q}) \bar\rho_{\vec q}^\dagger \bar \rho_{\vec q}
\end{equation}
where $\bar{\rho}_{\vec q} = \sum_je^{-|\vec q|^2\ell_B^2/4} e^{-i\vec q\cdot \vec R_{j}}$ is the LL-projected density operator in terms of particle guiding center coordinates $\vec R_j$, we've dropped the normal ordering for convenience, and $A$ is system size. It is helpful to distinguish two levels of approximation: (1) the variational approximation and (2) the variational \textit{and} perturbative ($V_{\vec G}\ll$ Coulomb) approximation. Both approximations are in the band-projected limit where the cyclotron gap $\hbar \omega_c$ is the largest energy scale. 

\par 
\textit{(1) Variational.--- } Recall that our variational ansatz is 
\begin{equation}
  \ket{\phi^\alpha_{\vec{q}}} =  \sum_{\vec{G}} c^\alpha_{\vec{q}+\vec{G}} \bar{\rho}_{\vec{q} + \vec{G}} \ket{\rm 0}
\end{equation} 
where $\vec q\in$ BZ and $\vec{G}$ is a reciprocal lattice vector, which has energy
\begin{equation}\label{eq:withoutparity}
      E^\alpha_{\vec{q}} -E_0 = \frac{\sum_{\vec{G}_1 \vec{G}_2} c^{\alpha \ast}_{\vec{q}+\vec{G}_1} c^\alpha_{\vec{q}+\vec{G}_2} \langle{0}|\bar{\rho}^{\dagger}_{\vec{q} + \vec{G}_1} [H, \bar{\rho}_{\vec{q} + \vec{G}_2}] |{0}\rangle}{\sum_{\vec{G}_1\vec{G}_2} c^{\alpha\ast}_{\vec{q}+\vec{G}_1}c^\alpha_{\vec{q}+\vec{G}_2} \bar{s}(\vec q,\vec G_1,\vec G_2)}
\end{equation} 
where $(H-E_0)\ket{0} =0$ and $\bar{s}(\vec q,\vec G_1,\vec G_2)= \langle 0|\bar{\rho}^\dagger_{\vec{q}+\vec{G}_1} \bar{\rho}_{\vec{q}+\vec{G}_2}  |0   \rangle$ is the projected static structure factor. If we also assume that $c_{\vec q+G}^{\alpha\ast} = c_{-\vec q-G}^{\alpha}$, a constraint on our ansatz which is natural in the presence of parity symmetry, then the above can be written as 
\begin{equation}\label{eq:withparity}
      E^\alpha_{\vec{q}} -E_0 = \frac12 \frac{\sum_{\vec{G}_1 \vec{G}_2} c^{\alpha \ast}_{\vec{q}+\vec{G}_1} c^\alpha_{\vec{q}+\vec{G}_2} \langle{0}|[\bar{\rho}^{\dagger}_{\vec{q} + \vec{G}_1}, [H, \bar{\rho}_{\vec{q} + \vec{G}_2}]] |{0}\rangle}{\sum_{\vec{G}_1\vec{G}_2} c^{\alpha\ast}_{\vec{q}+\vec{G}_1}c^\alpha_{\vec{q}+\vec{G}_2} \bar{s}(\vec q,\vec G_1,\vec G_2)}.
\end{equation} 
Using the Girvin-Macdonald-Platzman algebra~\cite{girvin1986magneto} 
\begin{equation}\label{eq:S8}
    [\bar\rho_{\vec q_1},\bar\rho_{\vec q_2}] = f_{\vec q_1}^{\vec q_2} \bar\rho_{\vec q_1+\vec q_2},\qquad f_{\vec q_1}^{\vec q_2} = 2ie^{\vec q_1\cdot \vec q_2\ell_B^2/2}\sin(\vec q_1\wedge \vec q_2 \ell_B^2/2),
\end{equation}
we can simplify the variational energies. It is useful to collect the following expressions: 
\begin{equation}\label{eq:someformulas}
    \begin{aligned}
       &[H,\bar\rho_{\vec q+\vec G_2}] = \sum_{\vec G} V_{\vec G} f_{\vec G}^{\vec q+\vec G_2} \bar\rho_{\vec q+\vec G+\vec G_2} + \frac{1}{2A} \sum_{\vec q'} U(\vec q') f_{\vec q'}^{\vec q+\vec G_2} \{\bar\rho_{\vec q+\vec q'+\vec G_2} ,\bar\rho_{-\vec q'}\}\\
        &\frac12 [\bar\rho_{-\vec q-\vec G_1},[H,\bar\rho_{\vec q+\vec G_2}]] = \\
        &\frac12\sum_{\vec G} V_{\vec G} f_{\vec G}^{\vec q+\vec G_2} f_{-\vec q-\vec G_1}^{\vec q+\vec G+\vec G_2} \bar\rho_{-\vec G_1+\vec G+\vec G_2} +\frac{1}{4A} \sum_{\vec q'} U(\vec q') f_{\vec q'}^{\vec q+\vec G_2} \left(f_{-\vec q-\vec G_1}^{-\vec q'} \{\bar\rho_{-\vec q-\vec q'-\vec G_1},\bar\rho_{\vec q+\vec q'+\vec G_2}\} +f_{-\vec q-\vec G_1}^{\vec q+\vec q'+\vec G_2}\{\bar\rho_{\vec q'+\vec G_2-\vec G_1},\bar\rho_{-\vec q'}\}\right)
    \end{aligned}
\end{equation}
where we've assumed the interaction satisfies $U(\vec q)=U(-\vec q')$ and $\{A,B\} = AB+BA$. Altogether, the energy without assuming parity symmetry (Eq.~\eqref{eq:withoutparity}) becomes 
\begin{equation}\label{eq:withoutparity2}
    E_{\vec q}^{\alpha} - E_0 = \frac{\sum_{\vec G_1\vec G_2}c_{\vec q+\vec G_1}^{\alpha\ast} c_{\vec q+\vec G_2}^\alpha \left(\sum_{\vec G}V_{\vec G}f_{\vec G}^{\vec q+\vec G_2} \bar s(\vec q,\vec G_1,\vec G+\vec G_2) + \frac{1}{2A} \sum_{\vec q'} U(\vec q')f_{\vec q'}^{\vec q+\vec G_2} \langle 0 |\bar\rho_{-\vec q-\vec G_1}\{\bar\rho_{\vec q+\vec q'+\vec G_2},\bar\rho_{-\vec q'}\}|0\rangle\right)}{\sum_{\vec G_1\vec G_2}c_{\vec q+\vec G_1}^{\alpha\ast} c_{\vec q+\vec G_2}^\alpha \bar s(\vec q,\vec G_1,\vec G_2)}
\end{equation}
and the energy assuming parity (Eq.~\eqref{eq:withparity}) becomes
\begin{equation}\label{eq:withparity2}
    E_{\vec q}^{\alpha} - E_0 = \frac{\sum_{\vec G_1\vec G_2}c_{\vec q+\vec G_1}^{\alpha\ast} c_{\vec q+\vec G_2}^\alpha \left(\frac12 \sum_{\vec G} V_{\vec G} f_{\vec G}^{\vec q+\vec G_2} f_{-\vec q-\vec G_1}^{\vec q+\vec G + \vec G_2} \langle0|\bar\rho_{-\vec G_1+\vec G_2+\vec G}|0\rangle + \frac{1}{4A}\sum_{\vec q'}U(\vec q') f_{\vec q'}^{\vec q+\vec G_2}   T^{\vec q,\vec q'}_{\vec G_1,\vec G_2}\right)}{\sum_{\vec G_1\vec G_2}c_{\vec q+\vec G_1}^{\alpha\ast} c_{\vec q+\vec G_2}^\alpha \bar s(\vec q,\vec G_1,\vec G_2)}
\end{equation}
where 
\begin{equation}
    T^{\vec q,\vec q'}_{\vec G_1,\vec G_2} =f_{-\vec q-\vec G_1}^{-\vec q'} (\bar s(\vec q+\vec q',\vec G_1,\vec G_2) +\bar s(-\vec q-\vec q',-\vec G_2,-\vec G_1) ) + f_{-\vec q-\vec G_1}^{\vec q+ \vec q'+ \vec G_2} (\bar s(-\vec q',\vec G_{12},0)+\bar s(\vec q',0,\vec G_{21})).
\end{equation}
We remark that Eq.~\eqref{eq:withoutparity2} involves the structure factor and a three-density expectation value, while Eq.~\eqref{eq:withparity2} only involves the structure factor and single-density expectation value. Naturally, the latter is easier to compute, but either may be computed numerically in principle for finite-size systems. \par 
\textit{(2) Variational + perturbative.--- } The perturbative variational calculation assumes $V$ is small, where
\begin{equation}
    H = H_C + V,\quad H_C =  \frac{1}{2A} \sum_{\vec{q}} U(\vec{q}) \bar\rho_{\vec q}^\dagger \bar \rho_{\vec q},\quad V = \sum_{\vec{G}}  V_{\vec{G}}\bar{\rho}_{\vec G}.
\end{equation}
Let us denote the eigenstates and energies of the Coulomb Hamiltonian $\hat H_C$ by $\ket{\Psi_{\vec k,m}^{(0)}}$ and $E_{\vec k,m}^{(0)}$, where $\vec k$ denotes (unrestricted) many-body momentum and $m$ distinguishes states with the same $\vec k$. We assume that the Coulomb ground state is a translationally invariant liquid (valid for FQH states) and denote it $\ket{\Psi_{0}^{(0)}}$. We also label the unperturbed energies in the crystal Brillouin zone as band energies by $E_{\vec k,m}^{(0)}$, with the (unperturbed) ground state energy denoted by $E_0^{(0)}$. We consider states of the form
\begin{equation}
 \ket{\phi_{\vec k}^{(0)}}  = \frac{1}{\sqrt{\bar s_0(\vec k)}} \bar\rho_{\vec k} \ket{\Psi_0^{(0)}}
\end{equation}
where $\bar s_0(\vec k) =  \langle \Psi_0^{(0)}| \bar\rho_{-\vec k} \bar\rho_{\vec k}|\Psi_{0}^{(0)}\rangle$. In the variational + perturbative approach, for any $\vec k$, we diagonalize $H$ in the subspace spanned by the $\ket{\phi_{\vec k+\vec G}^{(0)}}$ to leading order in $V$. We give two examples for high-symmetry points below. \par 
\textit{$\Gamma$ excitation.--- } We project $V$ into the manifold of 6 SMA states which are connected via $V$ to $\Gamma$: 
\begin{equation}
\begin{aligned}
\langle \phi_{\vec G_i}^{(0)}|   \hat V |\phi_{\vec G_j}^{(0)}\rangle &= V_{\vec G_i-\vec G_j} \langle \phi_{\vec G_i}^{(0)} |\bar \rho_{\vec G_i-\vec G_j}|\phi_{\vec G_j}^{(0)}\rangle\\
 &= \frac{V_{\vec G_i-\vec G_j}}{N\bar s_0(\vec G)} \langle0|\bar \rho_{-\vec G_i}\bar \rho_{\vec G_i - \vec G_j} \bar \rho_{\vec G_j}|0\rangle 
 \end{aligned}
\end{equation}
where we denote $\bar s_0(\vec G_i)$, which are equal for all $i$, by $\bar s_0(\vec G)$. We assume
\begin{equation}
    V_{\vec G_1} = V_{\vec G_3} =V_{\vec G_5} =V_{\vec G_2}^* = V_{\vec G_4}^* = V_{\vec G_6}^* = -V_0
\end{equation}
and we define
\begin{equation}\label{eq:s30}
    S_3 = \langle 0|\bar\rho_{\vec G_1} \bar \rho_{\vec G_3} \bar \rho_{\vec G_5}|0\rangle. 
\end{equation}
Then the Hamiltonian in this subspace can be written
\begin{equation}
\begin{aligned}
 H\left.\right|_{\Gamma} =   \Delta_{\vec G}- \frac{1}{\bar s_0(\vec G)} \begin{pmatrix}
        0  & V_0^*S_3 & 0&0&0&V_0^*S_3^*\\
        V_0S_3^* & 0 & V_0S_3 &0 & 0 &0\\
        0 & V_0^*S_3^* & 0 & V_0^*S_3 & 0 & 0\\
        0 & 0 &V_0S_3^* & 0 & V_0S_3 & 0\\
        0 & 0 & 0 & V_0^*S_3^* & 0 & V_0^*S_3\\
        V_0S_3& 0 & 0 & 0 & V_0S_3^* &0 
    \end{pmatrix}
\end{aligned}
\end{equation}
where $\Delta_{\vec G} = \langle \Psi_{\vec G_i,0}^{(0)}|H_C - E_0^{(0)}|\Psi_{\vec G_i,0}^{(0)}\rangle$ (for all $i$) is the exact Coulomb energy of the lowest excitation at $\vec G_i$. We introduce $C_3$-angular momentum eigenstates parity-related 
\begin{equation}
\begin{aligned}
\ket{\ell,1} &= \frac{1}{\sqrt{3}} (\ket{\phi_{\vec G_1}^{(0)}} + e^{2\pi i \ell/3} \ket{\phi_{\vec G_3}^{(0)}} + e^{4\pi i\ell/3} \ket{\phi_{\vec G_5}^{(0)}})\\
\ket{\ell,2} &= \frac{1}{\sqrt{3}} (\ket{\phi_{\vec G_2}^{(0)}} + e^{2\pi i \ell/3} \ket{\phi_{\vec G_4}^{(0)}} + e^{4\pi i\ell/3} \ket{\phi_{\vec G_6}^{(0)}})
\end{aligned}
\end{equation}
for $\ell=-1,0,1$. The lowest quasielectron-quasihole state is a linear combination of the above states with angular momentum $\ell = \argmax_\ell[\cos(\psi + \ell\pi/3)]$ and energy
\begin{equation}
    E_0^{(1)} =E_0^{(0)} + \Delta_{\vec G} -2\frac{|V_0\eta|}{\bar s_0(\vec G)} \max_{\ell}[\cos(\psi + \ell\pi/3)],
\end{equation}
where $S_3 = |S_3|e^{i\psi}$ (the phase of $V_0$ does not enter). \par 

\textit{$K$-roton.--- } The points $K_1,K_2,K_3$ satisfy $K_1-K_2 = \vec G_2, K_2-K_3 = \vec G_4, K_3-K_1 = \vec G_6$. The Hamiltonian in this subspace is 
\begin{equation}
\begin{aligned}
    H\left.\right|_K &= \Delta_K + \frac{1}{\bar s_0(K)} \begin{pmatrix}
         0 & V_{K_1-K_2}\langle 0|\bar\rho_{-K_1} \bar\rho_{K_1-K_2}\bar\rho_{K_2}|0\rangle &V_{K_1-K_3} \langle 0|\bar\rho_{-K_1} \bar\rho_{K_1-K_3}\bar\rho_{K_3}|0\rangle\\ V_{K_2-K_1}\langle 0|\bar\rho_{-K_2} \bar\rho_{K_2-K_1}\bar\rho_{K_1}|0\rangle & 0 & V_{K_2-K_3} \langle 0|\bar\rho_{-K_2} \bar\rho_{K_2-K_3}\bar\rho_{K_3}|0\rangle \\ V_{K_3-K_1} \langle 0|\bar\rho_{-K_3} \bar\rho_{K_3-K_1}\bar\rho_{K_1}|0\rangle & V_{K_3-K_2} \langle 0|\bar\rho_{-K_3} \bar\rho_{K_3-K_2}\bar\rho_{K_2}|0\rangle & 0 
    \end{pmatrix}\\
    &= \Delta_K - \frac{1}{\bar s_0(K)}\begin{pmatrix}
         0 & V_0^*S_3 &V_0 S_3^* \\ V_0S_3^* & 0 & V_0^* S_3 \\ V_0^* S_3 & V_0 S_3^* & 0 
    \end{pmatrix}
\end{aligned}
\end{equation}
where we defined $\bar s_0(K) \equiv \bar s_0(K_i)$ (for $i=1,2,3$) and now
\begin{equation}
    S_3 = \langle 0 |\bar\rho_{-K_1} \bar\rho_{K_2-K_1} \bar\rho_{K_2}|0\rangle
\end{equation}
(not to be confused with Eq.~\eqref{eq:s30}). We introduce the $C_3$-angular momentum eigenstate
\begin{equation}
    \ket{\ell} = \frac{1}{\sqrt{3}}\sum_{j=0}^2 e^{2\pi ij\ell/3}\ket{\phi_{K_j}^{(0)}}.
\end{equation}
The lowest $K$-roton is the angular momentum eigenstate with $\ell = \argmax_\ell[\cos(\psi + 2\pi \ell/3)]$ and energy
\begin{equation}
    E_K^{(0)} = E_0^{(0)} + \Delta_K - 2 \frac{|V_0S_3|}{\bar s_0(K)} \max_\ell[\cos(\psi+2\pi \ell/3)]
\end{equation}
where $\Delta_K = \langle \Psi_{K,0}^{(0)}|H_C-E_0^{(0)}|\Psi_{K,0}^{(0)}\rangle$ and $V_0S_3 = |V_0S_3| e^{i\psi}$. Note that unlike the case for the quasielectron-quasihole state, the phase of $V_0$ enters $E_K^{(0)}$. 

\section{Low-energy spectral weight in ordinary Landau levels}\label{sec:SM_LLweight}

It is a well-known corollary of Kohn's theorem that the low-energy optical spectral weight of an ordinary Landau level system vanishes. For completeness, we review a derivation here, originally due to GMP~\cite{girvin1986magneto}. We have 
\begin{equation}
\begin{split}
    W &= \frac{\pi e^2 }{2A\hbar^2} \lim_{q\to 0} \frac{1}{q^2} \langle 0 | [ \bar{\rho}_{\vec q}^{\dagger},[\bar{H}, \bar{\rho}_{\vec q} ]]| 0\rangle\\
    &\equiv \frac{\pi e^2n }{\hbar^2} \lim_{q\to 0} \frac{1}{q^2} F(\vec q).
\end{split}
\end{equation}
In an ordinary Landau level, the projected Hamiltonian has only the interaction piece, $\bar H = \frac{1}{2A} \sum_{\vec q'} U(\vec q') \bar\rho_{\vec q'}^\dagger \bar\rho_{\vec q'}$. Then $F(\vec q)$ may be evaluated using the GMP algebra Eq.~\eqref{eq:S8}. We quote the result from Eq.~\eqref{eq:someformulas} (we use $\bar s(\vec q) = \bar s(-\vec q)$ in the parity-symmetric Landau level): 
\begin{equation}
\begin{aligned}
    F(\vec q) &= \frac{1}{4A} \sum_{\vec q'} U(\vec q') f_{\vec q'}^{\vec q} \left(f_{-\vec q}^{-\vec q'} \langle 0|\{\bar\rho_{-\vec q-\vec q'},\bar\rho_{\vec q+\vec q'}\}|0\rangle + f_{-\vec q}^{\vec q+\vec q'} \langle 0|\{\bar\rho_{\vec q'},\bar\rho_{-\vec q'}\}|0\rangle \right) \\
    &= \frac{1}{2A} \sum_{\vec q'} U(\vec q') f_{\vec q'}^{\vec q} \left(f_{\vec q}^{\vec q'} \bar s(\vec q+\vec q') + f_{-\vec q}^{\vec q+\vec q'} \bar s(\vec q')\right)
\end{aligned}
\end{equation}
where $f_{\vec q}^{\vec q'} = 2ie^{\vec q\cdot \vec q'\ell_B^2/2}\sin(\vec q\wedge \vec q' \ell_B^2/2)$ and $\bar s(\vec q)$ is the projected structure factor. We expand at small $\vec q$ to obtain 
\begin{equation}
\begin{aligned}
    F(\vec q) &=\frac{1}{2A} (2i)^2\sum_{\vec q'} U(\vec q') (\vec q'\wedge \vec q\ell_B^2/2)^2\left(\bar s(\vec q+\vec q')-\bar s(\vec q')\right) +\cdots 
\end{aligned}
\end{equation}
where subleading terms are higher order in $|q|$. Next, using parity symmetry, it is possible to show that the leading term above vanishes, and hence that $F(\vec q)$ vanishes to at least order $q^4$. It follows that $W$ vanishes identically.

\section{Low-energy spectral weight for modulated Landau levels}\label{sec:SM_modLLweight}

In this section, we derive the formula for the low-energy sum rule for modulated Landau levels using another approach based on projected current operators that gives the same final answer. It is first instructive to review the general formalism for projected observables before applying it to a Chern band (or Landau level).\par 

\textit{Projection: general formalism.--- }Assume our Hamiltonian can be written as \begin{equation}
    H = T+V,\qquad T= \sum_n \varepsilon_n P_n
\end{equation}
where $\varepsilon_n$ are unperturbed energies, $P_n$ is a projection onto the $n$'th energy eigenspace, and $|V| \ll |\varepsilon_n-\varepsilon_m|$ for all $n,m$. Then one may wish to consider the projected Hamiltonian and projected observables in the $n$'th eigenspace. \par 
Let's focus on the projected Hamiltonian first. Throughout we use the overbar to denote operators defined only on the projected Hilbert space. We wish to find an operator $\bar H_n$ such that if $H\ket{\psi_i} = E_i \ket{\psi_i}$ is a solution to the full Schr\"odinger equation, where $E_i = \varepsilon_n + O(V)$, then $\bar H_n \ket{\tilde \psi_i} = E_i \ket{\bar \psi_i}$, where $\ket{\bar \psi_i} = P_n\ket{\bar \psi_i}$. The solution to leading order is 
\begin{equation}\label{eq:Hneff}
\bar H_n = \varepsilon_n + P_n V P_n + \sum_{m \neq n} \frac{1}{\varepsilon_n-\varepsilon_m} P_n V P_m V P_n + \mathcal{O}(V^3).
\end{equation}
Next, given some operator $M$ and two not necessarily distinct eigenvectors of $H$ denoted $\ket{\psi_1}$, $\ket{\psi_2}$ in the \textit{same} manifold of states near $E_n$, we want to find the operator $\bar M$, defined only on the projected Hilbert space, such that 
\begin{equation}
    \bra{\psi_1} M \ket{\psi_2} = \bra{\bar{\psi}_1} \bar M \ket{\bar{\psi}_2}, \qquad \ket{\bar{\psi}_i} = P_n \ket{\psi_i}.
    \label{eq:tm2}
\end{equation}
The solution to leading order is
\begin{equation}
    \bar M = P_n M P_n + \sum_{m \neq n} \frac{ 
    P_n  M P_m V P_n - P_n  V P_m M P_n 
    }{\varepsilon_n - \varepsilon_m } + \mathcal{O}(V^2).
    \label{eq:tm}
\end{equation}
Finally, given some operator $M$ and two distinct eigenvectors of $H$ denoted $\ket{\psi_1}$, $\ket{\psi_2}$ in \textit{different} manifolds of states near $E_n$ and $E_m$ respectively ($n\neq m$), we want to find the operator $\tM$ satisfying 
\begin{equation}
    \braket{\psi_1}{M|\psi_2} = \braket{\bar \psi_1 }{\bar M |\bar \psi_2},\qquad \ket{\bar \psi_1} = P_n \ket{\psi_1},\quad \ket{\bar \psi_2} = P_m \ket{\psi_2}. 
\end{equation}
The solution is
\begin{equation}
    \bar M = P_n M P_m + \sum_{\ell \neq m} \frac{P_nMP_\ell VP_m}{\varepsilon_m-\varepsilon_\ell} + \sum_{\ell\neq n}\frac{P_n VP_\ell MP_m}{\varepsilon_n-\varepsilon_\ell} + \mathcal{O}(V^2).
\end{equation}

\textit{Landau level projected theory.--- }  Quite generally, we consider electrons in a magnetic field and both 1-body and 2-body potentials:
\begin{subequations}
    \begin{align}\label{eq:cont}
        H  &= \sum_i H_0(\p_i-e\A) + V(\r_i) + \frac12\sum_{i\neq j}U(\r_i-\r_j)\\
        &= H_0 + H_C
    \end{align}
\end{subequations}
where $H_0(\p) = p^2/2m$ and $\A$ describes a uniform magnetic field $B\hat z$. If $V = 0$ and $U$ preserves Galilean invariance, then Kohn's theorem implies 
\begin{equation}
    \sigma_{\alpha\alpha}(\omega) = \frac{\pi e^2 n}{2m} \left(\delta(\omega-\omega_c) + \delta(\omega+\omega_c)\right).
\end{equation}
Again, we'll be interested in the low-energy spectral weight for when Galilean invariance is broken and $\Lambda$ is intermediate between the strengths of $V$ and $\hbar\omega_c$. With Galilean invariance, $W$ vanishes. Turning on $V$ breaks Galilean invariance and $W$ could be nonvanishing.
\par 
For $V,U\ll \omega_c$, we can project the current density operator to a Landau level. The current density operator is
\begin{equation}
    \begin{aligned}
    \vec{j} =  -\frac{e (\vec{p}+e\vec A)}{mA}
    & = \frac{i e \omega_c }{A}\left( a^\dagger \vec{d}^* - a \vec{d} \right)
    \end{aligned}
\end{equation}
where $\vec A$ is the gauge field and $A$ is the area. Moreover, $\vec{d} =  (\ell_B, i \ell_B)^T/\sqrt 2$. Consequently, the projected current density operator is
\begin{equation}
    \begin{aligned}
    \vec{\bar{j}} & = \frac{1}{\hbar\omega_c} 
    P_n \vec{j} ( P_{n-1} - P_{n+1}) H_C P_n  + \mathrm{h.c.} 
     \\
    & = - \frac{i e}{\hbar A} \left(
    P_n a^\dagger \vec{d}^* P_{n-1} H_C P_n  + P_n a \vec{d} P_{n+1} H_C P_n \right)  + \mathrm{h.c.}
    \\
    & =\frac{i e}{\hbar A} [ \vec{R} , \bar{H}_C ] 
     \end{aligned}
\end{equation}
where subleading terms are dropped and $\vec{R} = \sum_i^N \vec{R}_{i}$ is the guiding center coordinate. Since our focus is FCIs, we assume the ground state is insulating; hence the Drude weight vanishes and we can write $W = \int d\omega \, \text{Re } \sigma_{aa}^L(\omega)$ as follows~\cite{Resta2018Sep}
\begin{equation}
    \begin{aligned}
        W &= \frac{\pi A}{\hbar}\sum_{n\neq 0} \frac{\langle 0|j_a|n\rangle \langle n |j_a|0\rangle}{\omega_{n0}}\\
        &=\frac{\pi A}{\hbar}\sum_{n\neq 0} \frac{\langle \bar 0|\bar j_a|\bar n\rangle \langle \bar n |\bar j_a|\bar 0\rangle}{\omega_{n0}}\\
        &= \frac{\pi e^2}{2\hbar^2 A} \langle \bar 0 |[ R_a,[\bar H_C, R_a]]|\bar 0\rangle\\
        &= \frac{\pi e^2}{2\hbar^2 A}\lim_{q\to 0} \frac{1}{q^2} \langle \bar 0 |[\bar \rho_{\vec q}^\dagger ,[\bar H_C,\bar \rho_{\vec q}]]|\bar 0\rangle 
    \end{aligned}
\end{equation}
where $\vec q = q \hat a$. This is in agreement with results quoted in the main text. This is in general nonvanishing (as the next section confirms). Although finite intraband optical spectral weight is expected once continuous translation symmetry is broken, the perturbed Landau level clarifies that optical darkness is not a generic property of FCIs, but instead relies on additional assumptions. 

\section{Optical conductivity in periodically modulated Landau levels}\label{sec:SM_modLLoptical}
\begin{figure}[t!]
    \centering
    \includegraphics[width=0.7 \linewidth]{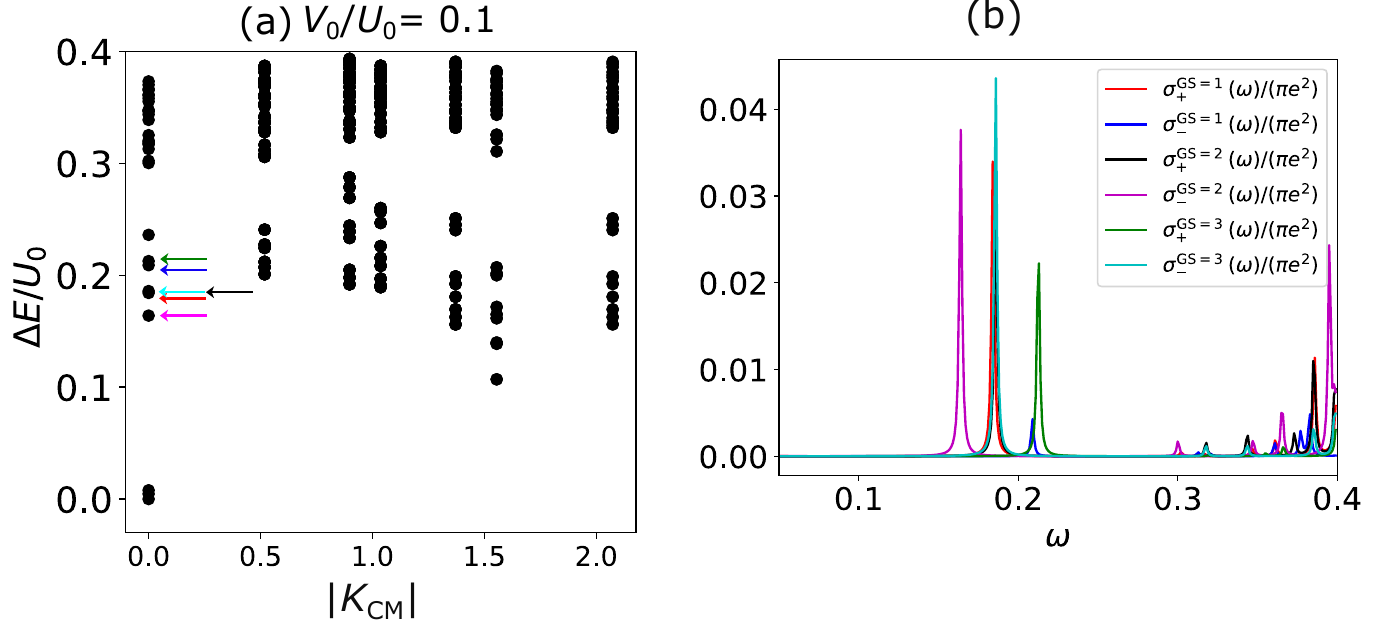}
    \caption{(a) The many-body spectrum at $\nu = 1/3$ for the Hamiltonian \eqref{eq:suppLLprojected} as a function of $|\vec{K}_{\rm CM}|$. (b) Left/right circularly polarized optical conductivity $\sigma^{\rm GS}_{\pm}(\omega)$ calculated  for the three quasi-degenerate FCI ground states. These show low-lying %
    quasielectron-quasihole states which are highlighted in panel (a). An artificial broadening $\eta = 0.01  \> U_0$ is introduced for clarity and $\hbar = 1$. Calculations were done on a 27 site cluster with equal aspect ratio similar to Fig. \ref{figS1}(b). $U_0 = e^2 /( 4 \pi \epsilon a_0)$ is interaction strength.}
    \label{figS2}
\end{figure}

Due to Kohn's theorem, all intra-LL excitations of any state in Landau levels are optically inactive and cannot be probed by optical absorption due to Galilean invariance. To go beyond this, we consider perturbing the Landau level system by adding a periodic potential. As discussed in the main text, the Hamiltonian projected to the lowest Landau level is given by
    \begin{equation}
\label{eq:suppLLprojected}
    H = \sum_{\vec{G}}  V_{\vec{G}}\bar{\rho}_{\vec G} + \frac{1}{2A} \sum_{\vec{q}} U(\vec{q}) :\bar\rho_{\vec q}^\dagger \bar \rho_{\vec q}:
\end{equation}
We consider a $D_6$-symmetric potential with six harmonics ($V_{\vec{G}} = - V_0$ for $\vec{G}_i= \frac{4 \pi}{\sqrt{3} a_0} [\cos 2\pi (i-1)/6,\sin 2\pi (i-1)/6]$ for $ i = 1,\dots,6$) and a period $a_0$ chosen such that the unit cell encloses one flux quantum ($\sqrt{3}a_0^2 = 4 \pi \ell_B^2$). The periodic potential will broaden the LLL to a dispersive Chern band. Due to the periodic potential, continuous magnetic translation symmetry is broken to a discrete one and we can only label the many-body states with momentum quantum number $\vec{K}_{\rm CM} \in {\rm BZ_{Bloch}}$. As shown in the many-body spectrum at $\nu = 1/3$ in Fig. \ref{figS2}(a), the periodic potential splits the previously degenerate many-body states (Fig. \ref{figS1}(b)) but the system is still in the FCI phase. Above the many-body ground states at $\vec{K}_{\rm CM} = 0$, we observe low-lying %
quasielectron-quasihole states that hybridize and split in energy due to the periodic potential.  Next, we calculate the response of the system to  left/right circularly polarized light \begin{equation}
\label{eq:opticalconductivity}
    \sigma^{\rm GS}_{\pm}(\omega) = \pi A \sum_{n \neq {\rm GS}} \dfrac{|\langle {\rm GS} |j_{\pm}|n \rangle|^2}{E_{n} - E_{\rm GS}}[\delta(\omega - (E_n - E_{\rm GS})) + \delta(\omega + (E_n - E_{\rm GS}))] 
\end{equation}
where $j_{\pm} = (j_x \pm i j_y)/\sqrt{2}$ with the current density operator given by $\vec{j} = \frac{1}{ A} \sum_{j} i e [\vec{R}_j,V(\vec{R_j})]$ with $\vec{R}_j$ is the guiding center operator for the $j$-th particle and $V(\vec{R}) = \sum_{\vec{G}} V_{\vec{G}} e^{-|\vec G|^2\ell_B^2/4} e^{-i\vec G\cdot \vec R}$. $A$ is the area of the system. $\sigma^{\rm GS}_{\pm}(\omega)$ (and any response function) can be calculated efficiently using the Lanczos algorithm without the need to find all eigenstates of the system as we explain in a later section. 

In Fig. \ref{figS2}(b), we plot $\sigma^{\rm GS}_{\pm}(\omega)$ for the three quasi-degenerate ground states. We find the conductivity to exhibit peaks at frequencies that correspond to the low-lying %
quasielectron-quasihole states directly showing that they are optically active (in contrast to the FQH case without any periodic potential). 

In addition, we observe in Fig. \ref{figS2}(b) peaks in the higher frequency range ($0.3 <\omega< 0.4$) that correspond to chiral graviton excitations \cite{haldane_geometrical_2011,liou_chiral_2019} (the $\q \rightarrow 0$ of the magneto-roton branch, shown Fig. \ref{figS1}(a)). These show that some of the chiral graviton modes can be optically active when a periodic potential is added. Indeed with a $C_n$ symmetric potential, angular momentum is defined only modulo $n$ and in principle, dipole matrix elements can be non-zero. We leave the detailed study of the optical response of these modes to future work.

 \begin{figure}[t!]
     \centering
     \includegraphics[width=0.5\linewidth]{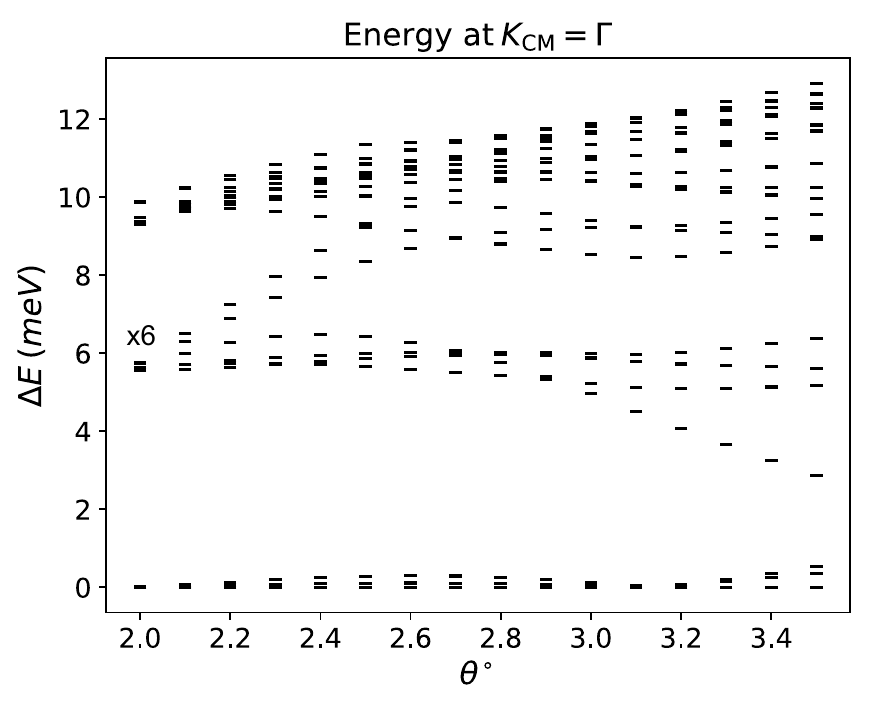}
    \caption{Many-body spectrum in the total momentum sector $K_{\rm CM} = 0$ as  a function of twist angle calculated for twisted ${\rm MoTe}_2$ at hole filling $n = 2/3$. There are 3 quasi-degenerate FCI ground states, above which, there exist low-lying collective modes which we argue are quasielectron-quasihole states based on the LL analogy.   }
     \label{fig:S3}
 \end{figure}

\section{Optical conductivity in the adiabatic model projected to the lowest Landau level}\label{sec:SM_adiabatic}

\begin{figure}
    \centering
    \includegraphics[width=0.6\linewidth]{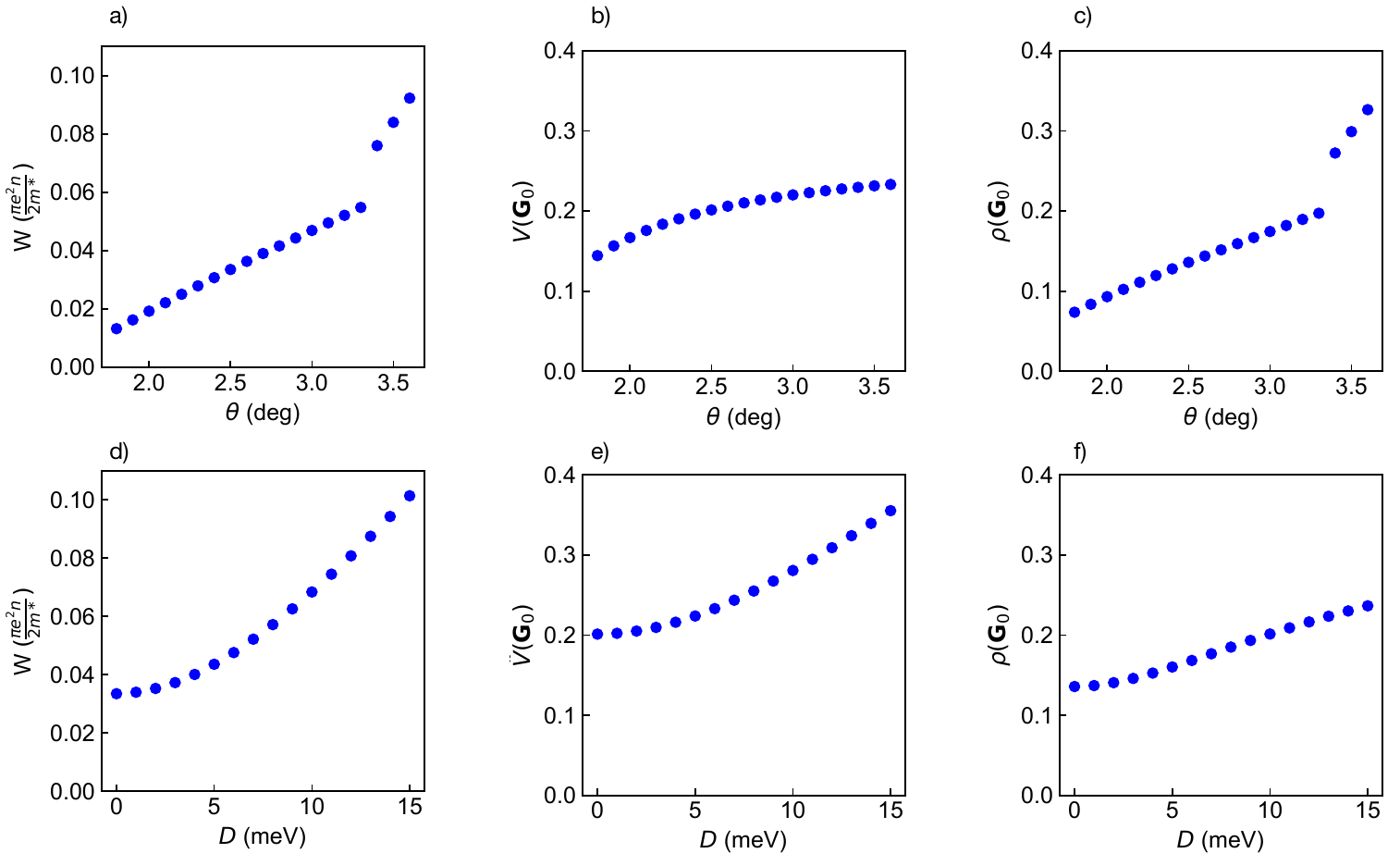}
    \caption{(a) Optical spectral weight in the adiabatic model for $t$MoTe$_2$ projected to the lowest Landau level as a function of twist angle calculated with Eq. \ref{eq:WpertLL_SM}. (b) Shows the lowest-harmonic Fourier component $V_{\bm G_0}$ of the effective projected potential and (c) shows the lowest Fourier component $\rho(\bm G_0)$ of the ground state charge density. The discontinuity in (a) and (c) is due to a level crossing among the three topologically degenerate ground states in the FCI phase.(d-f) Show the same quantities as a function of $D$ at $\theta=2.5^{\circ}.$ We use $\epsilon=10$ and the 27 site cluster.}
\label{fig:adiabatic_spectralweight}
\end{figure}

Here, we apply the equation
\begin{equation}\label{eq:WpertLL_SM}
    W = \frac{\pi e^2}{2\hbar^2 A}\sum_{\vec G}  \left(- V_{\vec G} \cdot \rho(\G)\right)(\hat q \wedge \vec G\ell_B^2)^2 
\end{equation}
(Eq.~\eqref{eq:WpertLL} of the main text) to compute the optical conductivity in the $n=\frac{2}{3}$ FCI phase of the adiabatic model of $t$MoTe$_2$ projected to the lowest Landau level \cite{morales2023magic}. We diagonalize the Hamiltonian Eq. \ref{eq:suppLLprojected} with the potential Fourier coefficients $V_{\bm G}$ determined from the adiabatic approximation to the continuum model of twisted MoTe$_2$. Concretely,
\begin{align}
    \begin{split}
        V_{\bm G} = \tilde V_{\bm G} + \frac{\hbar\omega_c}{2 B_0}\delta B_{\bm G}
    \end{split}
\end{align}
where $\tilde V_{\bm G}$ is the effective scalar potential and $\delta B_{\bm G}$ is the effective magnetic field fluctuation in the adiabatic approximation. Refs. \cite{morales2023magic} and \cite{reddy2024non} include further details. Unlike Ref. \cite{morales2023magic}, we exclude the $\delta \bm A^2$ term upon projecting to the lowest Landau Level because it is quadratic rather than linear in $\frac{\delta B_{\bm G}}{\ell_B |\bm G| B_0}$. We keep all Fourier components $\tilde V_{\bm G}$ rather than only the lowest harmonics as Ref. \cite{morales2023magic}. However, as argued in Ref. \cite{morales2023magic}, the lowest Fourier harmonic of $V_{\bm G}$ quantitatively dominates when projected to the lowest Landau level.
\par 

Fig. \ref{fig:adiabatic_spectralweight} shows the optical spectral weight thus obtained. At twist angles near $\theta = 3.5^{\circ}$ and / or interlayer potentials near $D=15$ meV, the intra-band optical spectral weight $W$ can reach nearly 10\% of the total optical spectral weight $\frac{\pi n e^2}{2m^*}$. Moreover, it is strongly tunable by displacement field. At $\theta=2.5^{\circ}$, while $W/\frac{\pi n e^2}{2m^*}$ is just below 4\% at $D=0$, it exceeds 10\% when $D=15$ meV. The optical spectral weight we obtain here is smaller than what we find for the original continuum model (see Fig.~\ref{fig:sumrule_TMDS} of the main text and Fig. \ref{fig:opticalweight_SM} in the SM). We believe that the lowest Landau level projection approximation is the cause of the reduction in optical spectral weight relative to the original continuum model.

\section{Quasielectron-quasihole excitations in twisted TMDs}\label{sec:SM_qeqh}

In this section, we study the low-lying collective modes of twisted ${\rm MoTe}_2$ in the many-body ground state. We use a 27 site cluster \cite{reddy2023fractional}
for which the quasi-degenerate FCI ground states at $n = 2/3$ occur at $\Gamma$ (total momentum $K_{\rm CM} = 0$). In Fig. \ref{fig:S3}, we show the many-body spectrum in this momentum sector as a function of twist angle. 

First, we focus on $\theta = 2.0^{\circ}$ where the system is the closest to the Landau-level limit \cite{morales2023magic}. At this \textit{magic angle}, we observe 6 low-lying states. Based on the Landau level analogy explained in Fig. \ref{figS1} for the 27-site cluster, there are six low-lying states in the FQH spectrum at two wavevectors $\vec{G}_1$ and $-\vec{G}_1$ that fold back to $\Gamma$. The observed states at $\theta = 2.0^{\circ}$ are in agreement with this physical picture, which gives strong evidence that they represent excitations consisting of a quasihole and quasielectron pair separated by a large distance. As a function of twist angle, the six quasielectron-quasihole excitations split in energy but, importantly, some of them remain isolated below the continuum at all twist angles. 

\section{Optical conductivity in twisted TMDs}\label{sec:SM_TMDoptical}

 \begin{figure}[t!]
      \centering
      \includegraphics[width=1.0\linewidth]{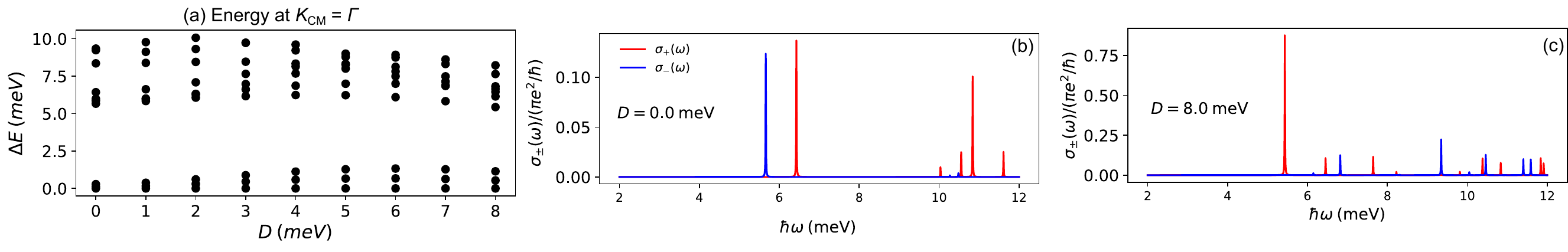}
      \caption{(a) Many-body energy spectrum of twisted ${\rm MoTe_2}$ at $n = 2/3$ at $\Gamma$ ($K_{\rm CM} = 0$ total momentum sector) as a function of displacement field $D$ at $\theta = 2.5^{\circ}$. There are three quasi-degenerate FCI ground states and low-lying collective modes. (b-c) Optical conductivity in the lowest many-body ground state for left ($\sigma^{+}$) and right ($\sigma^{-}$) circularly polarized light at (b) $D = 0.0$ and (c) $D = 8.0$ meV. An artificial broadening of $\eta = 0.01$ meV was used. }
    \label{fig:S4}
  \end{figure}
Having shown in the previous section that the low-lying collective modes of twisted ${\rm MoTe}_2$ are quasielectron-quasihole states, we now study the optical response of these modes. The optical conductivity for left/circularly polarized light is given by equation \eqref{eq:opticalconductivity} where the current density operator in our case is given by $j_{\alpha} =   \dfrac{e^2 \hbar k_{\alpha} }{m_* A} \sigma_{0} $ where ${\alpha = x,y}$ and $\sigma_{0}$ is identity matrix in layer space (see Eq.~\eqref{continuum} in the main text). 

In Fig. \ref{fig:S4}(b-c), we calculate the optical conductivity at $\theta = 2.5^{\circ}$ for displacement field $D = 0.0$ meV and $D = 8.0$ meV. We observe peaks at frequencies corresponding to the low-lying quasielectron-quasihole states (Fig. \ref{fig:S4}(a)). This provides a proof of principle that these modes are also optically active similar to the case of periodically modulated Landau levels. \par 
The previous work \cite{wolf2024intraband} suggested that intraband dipole transitions in $|C|=1$ ideal flat Chern bands are parametrically weak, and hence they are optically dark. Twisted MoTe$_2$ near the magic angle is a near-ideal flat Chern band with appreciable optical spectral weight, which is seemingly in tension with the results of \cite{wolf2024intraband}. It should be noted that \cite{wolf2024intraband} relied on stronger conditions than the ideal band condition, namely: (1) the LLL-like density matrix property of \cite{MacDonald1988Sep} and (2) that the charge density of the FCI is $1/m$ times the charge density of the filled flat band. We believe the source of discrepancy is due to these nonstandard assumptions, which need not apply here. A simple example illustrating the insufficiency of these assumptions is a flat ideal $C=1$ band with fluctuating quantum geometry, as may be realized by a Dirac electron in an inhomogeneous magnetic field. For sufficiently strong quantum geometry fluctuations, a Fermi liquid state results (as shown in Ref. \cite{Guerci2025Jun}), which has finite optical spectral weight.

\section{Low-frequency optical absorption in twisted TMDs}\label{sec:SM_TMDweight}

\begin{figure}[h]
    \centering
    \includegraphics[width=0.9\linewidth]{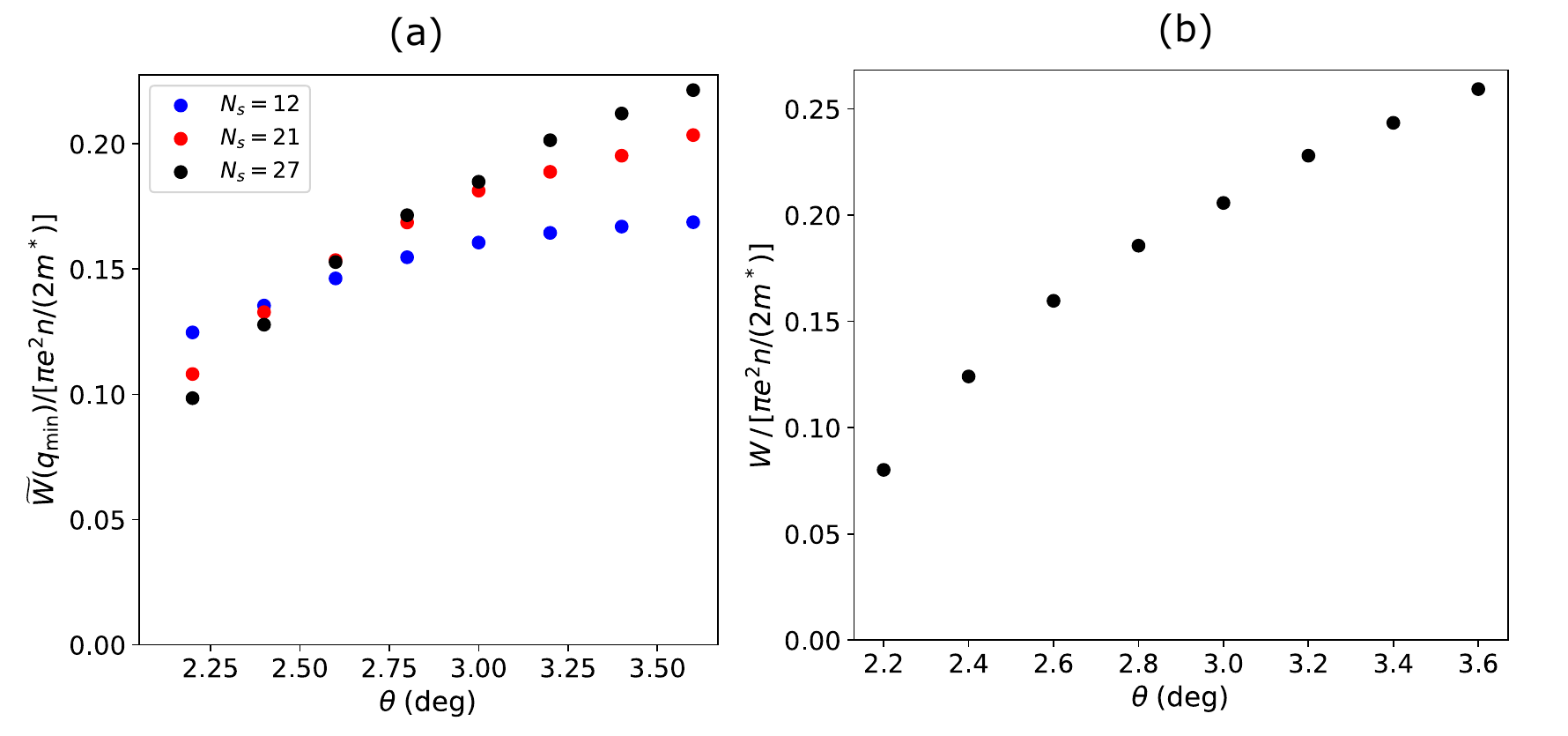}
    \caption{$\widetilde{W}(\q) = (\pi e^2 n F(\q))/(\hbar^2 q^2) $ computed at the smallest non-zero momentum $\q_{\rm min}$ as a function of twist angle. $N_s$ denotes the number of unit cells of the finite system. (b)  Intraband optical weight $W = \lim_{q \rightarrow 0} \widetilde{W}(\q) $, calculated by extrapolating  $\widetilde{W}(\q_{\rm min})$ with a linear fit, as a function of twist angle $\theta$. Calculations are done for twisted ${\rm MoTe}_2$  at $n= 2/3$. $W$ is given in units of $\pi e^2 n/(2m^*)$ where $n$ is the density and $m^*=0.62 m_e$ is the effective mass.}
    \label{fig:opticalweight_SM}
\end{figure}

In this section, we provide additional data for the low-frequency optical absorption $W$ (Eq.~\eqref{eq:spectralweight-band} in the main text) for twisted ${\rm MoTe}_2$ at different twist angles. In computing $W$, we first compute $\widetilde{W}(\q) = (\pi e^2 n F(\q))/(\hbar^2 q^2) $ at the smallest allowed momentum $\q_{\rm min}$ for three finite systems with $N_{s} = 12, 21, 27$ unit cells which all have $C_3$ symmetry (Fig. \ref{fig:opticalweight_SM} (a)). Then we fit $\widetilde{W}(\q_{\rm min})$ vs $\q_{\rm min}^2$ to a linear function and extract its intercept which estimates $W = \lim_{\q \rightarrow 0} \widetilde{W}(\q)$.  As shown in Fig.\ref{fig:opticalweight_SM}(b) , we find non-zero intraband optical absorption at various twist angles.  As discussed in the main text, the  intraband optical absorption spectrum is governed by the collective excitations (e.g quasielectron-quasihole pairs) of the FCI ground states

\section{Numerical Evaluation of Response Functions}\label{sec:SM_numerics}
In this section, we discuss efficient numerical methods to evaluate arbitrary response functions in frequency space.

\subsection{(a) Dynamical Response Functions}

Quite often, we are interested in evaluating a dynamical response function  \begin{equation}
\label{eq:I_omega}
I_B(\omega) = \sum_{n \neq 0} |\langle 0|B|n \rangle|^2 \delta(\omega - (E_n - E_0))
\end{equation} 
where $\ket{0}$ is the many-body state of interest (for example the ground state of a many-body Hamiltonian) with energy $E_0$ and $\ket{n}$  is the $n$-th eigenstate of the Hamiltonian with energy $E_n$. $B$ denotes the response operator (for example current or density). Let's define the Green's function $G_B(Z)$, \begin{equation}
\label{eq:greensfunction}
G_B(Z) = \langle 0|\dfrac{B^{\dagger}B}{Z-H}|0\rangle
\end{equation}
 Equation \eqref{eq:I_omega} can be recast in the following form
 \begin{equation}
 \label{eq:I_omega_2}
 I_B(\omega) = - \frac{1}{\pi} {\rm lim}_{\eta \rightarrow 0} \> {\rm Im} \>  G_B(\omega + E_0 + i \eta)
 \end{equation} where we have made use of the identity ${\rm lim}_{\eta \rightarrow 0} 1/ (\omega \pm i \eta)  = \mathcal{P} (1/\omega) \mp  2 \pi i \delta(\omega)$ with $\mathcal{P}$ denoting the principal value.  A finite $\eta$ corresponds to broadening the imaginary part from a delta function to a Lorentzian. To evaluate equation \eqref{eq:I_omega} or equation \eqref{eq:I_omega_2}, one can simply find all the eigenstates of the Hamiltonian. However, this is only possible for small systems. As shown in \cite{gagliano_dynamical_1987}, the dynamical correlation function \eqref{eq:I_omega_2} can be evaluated efficiently in the Krylov subspace of the normalized state $\ket{B} =  B\ket{0}/\sqrt{\langle 0|B^{\dagger} B |0 \rangle}$ defined as,
 
 \begin{equation}
 \label{eq:krylov}
 \mathcal{K}^{L}(B) = {\rm span}(\ket{B}, H\ket{B}, \dots, H^L\ket{B})
 \end{equation}
 This is the same idea behind the standard Lanczos algorithm for finding extremal eigenvalues of a matrix. The orthonormal basis $\ket{v_n}$ in $L$-th dimensional Krylov subspace $\mathcal{K}^L(B)$ is given by 
 \begin{equation}
 \label{eq:krylovbasis}
 \begin{split}
 & \ket{v_0} = \ket{B} \\
 & H \ket{v_n} = b_n \ket{v_{n-1}} + a_n \ket{v_n} + b_{n+1} \ket{v_{n+1}} \\
 & a_n = \langle v_n | H_n | v_n \rangle, \> b_{n+1} = || H \ket{v_n} - a_n \ket{v_n} - b_n \ket{v_{n-1}}||, \> b_0 = 0 \\
 \end{split}
 \end{equation}
 The Hamiltonian expressed in this basis takes the form of a tridiagonal matrix, 
 \begin{equation}
 \label{eq:Hrep}
  H = 
 \begin{pmatrix}
a_0  &  b_1 & 0  & 0  &   & 0  & 0 \\ b_1 & a_1 & b_2 & 0 & \cdots  & 0 & 0  \\ 
0 & b_2 & a_2  & b_3 & &  0 & 0 \\
0 & 0  & b_3 & a_3  & &  0 & 0 \\
  & \vdots & & & \ddots & & \vdots  \\
 0 & 0 & 0 & 0 & & a_{L-1} & b_L \\
 0 & 0 & 0 & 0 & \cdots & b_L & a_L 
 \end{pmatrix}
 \end{equation}
 In the Krylov subspace basis  \eqref{eq:krylovbasis}, the Green's function \eqref{eq:greensfunction} is given by 
 \begin{equation}
 G_B(Z) = \langle 0 | B^{\dagger} B |0 \rangle  [(Z-H)^{-1}]_{00}
 \end{equation}
 where $00$ denotes the element of the first row and first column. We then evaluate the inverse of  $(Z-H)^{-1}$ by partitioning it into a block $2\times2$ matrix, 
 \begin{equation}
 \begin{split}
 Z-H = \begin{pmatrix} 
 	Z-a_0 & (B^{(1)})^T \\
	B^{(1)}  &Z- H^{(1)}
	\end{pmatrix}\\
	[(Z-H)^{-1}]_{00} = (Z- a_0 - b_1^2 [(Z- H^{(1)})^{-1}]_{00})^{-1}
 \end{split} 
 \end{equation}
  Evaluating the inverse recursively, we arrive at a continued fraction expression for $G_B(Z)$, 
 \begin{equation} 
 \label{eq:contfrac}
 \begin{split}
 G_B(Z) = \langle 0 | B^{\dagger} B |0 \rangle  [(Z-H)^{-1}]_{00} \\
 = \dfrac{\langle 0 | B^{\dagger} B |0 \rangle}{Z- a_0 - \dfrac{b_1^2}{Z- a_1 - \dfrac{b_2^2}{Z-a_2 - \dots}}}
 \end{split}
 \end{equation}
 To evaluate $G_B(Z)$ in the many-body state $\ket{0}$, we first calculate $\ket{A} =  B\ket{0}/\sqrt{\langle 0|B^{\dagger} B |0 \rangle}$ then apply $L$ Lanczos iterations as given in \eqref{eq:krylovbasis} then we evaluate $G_B(Z)$ using equation \eqref{eq:contfrac}. The dynamical response function $I_A(\omega)$ can be calculated simply through equation \eqref{eq:I_omega_2}. The number of Lanczos iterations $L$ is increased until convergence is reached.  Therefore through this approach, we only need the Hamiltonian $H$, the state $\ket{0}$  and the response operator $B$. 
 \subsection{Example: Optical Conductivity}
 We apply this method to calculate the regular part of the optical conductivity which we copy below,
 
 \begin{equation}
    \sigma_{\pm}(\omega) = \frac{\pi A }{|\omega|} \sum_{n \neq 0} | \langle 0| j_{\pm} |n \rangle |^2[\delta(w - E_n + E_0) - \delta(w + E_n - E_0)]
\end{equation}
 
 With $j_{\pm}$  the current operator for left/right circular polarization  In this case, we take $B = j_{\pm}$ giving rise to 
 
\begin{equation}
\begin{split}
\sigma_{\pm}(\omega >  0) =  -  \frac{\pi A}{|\omega|}   \frac{1}{\pi} {\rm lim}_{\eta \rightarrow 0} \> {\rm Im} \>  G_{j_\pm}(\omega + E_0 + i \eta) \\
\sigma_{\pm}(\omega  < 0) =  -  \frac{\pi A}{|\omega|}   \frac{1}{\pi} {\rm lim}_{\eta \rightarrow 0} \> {\rm Im} \>  G_{j_\pm}(\omega - E_0 + i \eta)
\end{split}
\end{equation}
From $\sigma_{\pm}(\omega)$, one can readily compute the real part of longitudinal conductivity ${\rm Re} \sigma_{xx} $ or the imaginary part of the Hall conductivity  ${\rm Im} \sigma_{xy}$ by taking the sum or difference of $\sigma_{+}$ and $\sigma_{-}$.

\end{document}